\documentclass[doctor]{thesis}
\usepackage[cmex10]{amsmath}
\usepackage{amsthm,amssymb}

\usepackage[pdftex]{graphicx} % remove pdftex if you are not compiling to pdf
\graphicspath{{./figures/}} % this places all graphics in the figures subdirectory

\DeclareGraphicsExtensions{.pdf,.png,.jpg}

\usepackage[caption=false]{subfig}

\usepackage{booktabs}

\usepackage{url}
\usepackage{mathtools}
\usepackage{tikz}
\usepackage{xcolor}
\usepackage[textsize=small, textwidth=1.75cm]{todonotes}
\usetikzlibrary{calc}
\usetikzlibrary{shapes,snakes}
  \usetikzlibrary{positioning,arrows}
\tikzstyle{myarrows2}=[line width=1mm,draw=black,->] 
\tikzstyle{myarrows}=[line width=1mm,draw=blue,-triangle 45, postaction={draw, line width=1.5mm, shorten >=0.3mm, -}]
\tikzstyle{myarrows1}=[line width=1mm,draw=red,  dashed, postaction={draw, dashed, line width=1.1mm}]
\tikzstyle{myarrows3}=[line width=0.5mm, dashed, draw=black,-] 
\newcommand{\commentout}[1]{}

\usepackage{algorithm}
\tikzstyle{myarrows2}=[line width=1mm,draw=black,->] 
\tikzstyle{myarrows}=[line width=1mm,draw=blue, -triangle 60, 
 postaction={draw, line width=1.5mm, shorten <=0.3mm,-}]
\tikzstyle{myarrows1}=[line width=1mm,draw=red,  dashed, postaction={draw, dashed, line width=1.5mm}]
\tikzstyle{myarrows3}=[line width=0.5mm, dashed, draw=black,-]
\tikzstyle{myarrows4}=[line width=1mm,draw=red, -triangle 60, 
 postaction={draw, line width=1.5mm, shorten <=0.3mm,-}] 
\tikzstyle{myline1}=[line width=1mm,draw=blue,  dashed, postaction={draw, dashed, line width=1.5mm}]
\tikzstyle{myline2}=[line width=1mm,draw=red,  dashed, postaction={draw, dashed, line width=1.5mm}]

\usepackage{pdfpages}
\title{Analytical Cost Metrics : Days of Future Past}

\author{Nirmal Prajapati and Sanjay Rajopadhye}

\email{Nirmal.Prajapati@colostate.edu}

\department{Department of Computer Science}

\semester{Spring 2018}

\mycopyright{%
Copyright by Nirmal Prajapati\_\_ \\
All Rights Reserved
}

\abstract{%
}

\usepackage[pdfpagelabels,pdfusetitle,colorlinks=false,pdfborder={0 0 0}]{hyperref}

\begin{document} % preamble is complete, add any custom packages above
%%%%%%%%%%%%%%%%%%%%%%%%%%%%%%%%%%%%%%%%%%%%%%%%%%%%%%%%%%%%%%%%

\frontmatter % starts preliminary pages
%%%%%%%%%%%%%%%%%%%%%%%%%%%%%%%%%%%%%%%%%%%%%%%%%%%%%%%%%%%%%%%%

\maketitle              % insert title page
%\makemycopyright        % insert copyright page
%\makeabstract           % insert abstract page
%\makeacknowledgements   % insert acknowledgements page

% any extra preliminary pages can be added here
% below is an example of a dedication page
% the dedication page is optional
%\prelimtocentry{Dedication} % add table of contents entry
%\begin{flatcenter} % center without extra space
%
%    % page title
%    DEDICATION
%
%    %\vspace{3em} % place at top
%    \vfill % or center on page
%
%    \noindent \textit{I would like to dedicate this thesis to my dog fluffy.}
%    \vfill % fill extra space at bottom
%\end{flatcenter}
\newpage

%%\tableofcontents    % insert table of contents
%\listoftables       % insert list of tables (optional)
%\listoffigures      % insert list of figures (optional)

\mainmatter % starts thesis body
%%%%%%%%%%%%%%%%%%%%%%%%%%%%%%%%%%%%%%%%%%%%%%%%%%%%%%%%%%%%%%%%

% self notes
%\include{outline}

\chapter{Introduction}
\label{chap:introduction}
%%%%%%%%%%%%%%%%%%%%%%%%%%%%%%%%%%%%%%%%%%%%%%%%%%%%%%%%%%%%%%%%
%
%
As we move towards the exascale era, the new architectures must be capable of running the massive computational problems efficiently.  Scientists and researchers are continuously investing in tuning the performance of extreme-scale computational problems.  These problems arise in almost all areas of computing, ranging from big data analytics, artificial intelligence, search, machine learning, virtual/augmented reality, computer vision, image/signal processing to computational science and bioinformatics.  

With Moore's law driving the evolution of hardware platforms towards exascale, the dominant performance metric (time efficiency) has now expanded to also incorporate power/energy efficiency. Therefore the major challenge~\cite{gameover, kuck-hpc-challenges, exascale-bergman2008, exascale-shalf2010, exascale-dally2011power} that we face in computing systems research is: ``\emph{how to solve massive-scale computational problems in the most time/power/energy efficient manner}?''  

The architectures are constantly evolving making the current performance optimizing strategies less applicable and new strategies to be invented.  The solution is for the new architectures, new programming models, and applications to go forward together.  Doing this is, however, extremely hard.  There are too many design choices in too many dimensions.  The algorithms/applications may have a wide range of parameters, making them either intensely bandwidth bound or heavily compute bound, limited only by the hardware's operational throughput.  This significantly affects the algorithms used to obtain the best solution.  At the hardware end, the target platform affects not only the parallelization methods but also the tool-chains used.  To overcome this complications, cost models and simulators are used.

We propose the following strategy to solve the problem, i.e., \emph{``how to solve massive-scale computational problems in the most time/power/energy efficient manner}?'': 
\begin{enumerate}
\item \textbf{Models} Develop accurate analytical models (e.g. \emph{execution time, energy, silicon area}) to predict the cost of executing a given program.
\item \textbf{Complete System Design} Simultaneously optimize all the cost models for the programs (computational problems) to obtain the most \emph{time/area/power/energy efficient} solution. Such an optimization problem evokes the notion of codesign.
\end{enumerate}

Codesign---the simultaneous design of hardware and software---has two common interpretations~\cite{b1282de6ab8a4f1a8a962cb40310bc7f, gameover}.  \emph{System codesign}~\cite{exascale-DOE-2010, exascale-messina2017} is the problem of simultaneously designing hardware, run-time system, compilers, and programming environments of entire computing systems, typically in the context of large-scale, high-performance computing (HPC) systems and supercomputers. \emph{Application codesign}, also called \emph{hardware-software codesign}~\cite{codesign-thomas1993model, codesign-wolf1994hwsw-embedded, codesign-wolf2003decade}, is the problem of systematically and simultaneously designing a dedicated hardware platform and the software to execute a single application (program). The proposed approach is applicable to both contexts.

Solving the problem of \emph{System codesign} in its full generality is very difficult~\cite{exascale-DOE-2010, exascale-shalf2010, gameover, kuck-hpc-challenges} because of the wide range of (i) computational problems, (ii) the programming languages and abstractions used to express them, and (iii) varying target architectures, from data centers and cloud-based platforms, to distributed heterogeneous mobile platforms such as phones, and even ``things.''  Therefore, we suggest the following breakdown:
\begin{enumerate}
\item \textbf{Employ Domain-Specificity} Choose (i) a (small set/class) of programs, (ii) highly optimized hardware accelerators, and (iii) the optimal compiler transformations.
\item \textbf{Develop Cost Models} Develop accurate analytical cost (Time/Power/Area/Energy) models for performance prediction and optimization. 
\item \textbf{Mini Sub Prob - Specialize} Solve the codesign problem for the specific domain in (1) using the models in (2) i.e. optimize the program, compiler and hardware parameters simultaneously.
\item \textbf{Large Prob - Generalize} Repeat the above process to cover various important classes of programs, assign a weight to each class of programs and formulate an optimization problem for this weighted set of program parameters.  Simultaneously, solve for all the parameters and for all classes of programs.
\end{enumerate}
\section{Specialization}
\emph{Application codesign}~\cite{codesign-chiodo1994hwsw-embedded} is particularly important for embedded systems. In many uses of these systems (e.g., self-driving cars, computational fluid dynamics, neural networks, medical imaging, smart cameras, and  cyber-physical systems) general purpose platforms based on standard CPUs deliver inadequate ``performance'' on a combination of many cost metrics: speed/throughput, power/energy, weight, size, and manufacturing/fabrication cost, especially in volumes that the market can sustain.  As a result, specialized hardware is essential.

Hardware platforms for embedded systems are usually heterogeneous, with (instruction-set) programmable processors (CPUs and micro-controllers), accelerators that are either instruction-set programmable (e.g., GPUs), or ``hardware programmable'' ones like FPGAs and reconfigurable logic, as well as Application Specific Integrated Circuits (ASICs).  For HPC systems, ASICs are usually not a designer option.  In either case, the platforms have specialized, highly parallel, often fine-grain, components like FPGAs, GPUs, or DSPs, called \emph{accelerators}.  

The challenge is exacerbated when we consider the fact that accelerators are not one single architecture, and moreover, are constantly evolving.  For
example,
\begin{itemize}\itemsep 0mm
\item Google recently developed the \emph{\textbf{TPU}} (Tensor-flow Processing Unit), an ASIC to accelerate machine learning computations.  It is completely invisible to end users, who access it via the Tensor-Flow tool, and whose back end presumably makes direct library calls to TPUs.
\item Microsoft released \emph{\textbf{Catapult}}, an FPGA based fabric for accelerating large-scale data-center services.  It is a custom design, written in Verilog, and accessed via library calls.
\item At the other end of the spectrum, Facebook is developing its large scale machine learning applications using off-the-shelf GPUs and conventional tool chains: CUDA/OpenCL.
\end{itemize}
Mapping an application to an accelerator platform, \emph{even when the hardware is fixed}, is extremely difficult.  Codesign seeks to \emph{simultaneously design} the hardware itself, and is, therefore, an even harder problem.  Multiple cost metrics must be optimized, while still providing flexibility to the programmer and end user, and the design space is huge.

The key element of our approach is to exploit multiple forms of \emph{domain-specificity}~\cite{gameover}.  First, we tackle a specific (family of) computations that are nevertheless very important in many embedded systems.  This class of computations, called \emph{dense stencils}, includes the compute-intensive parts of many applications such as computational fluid dynamics, neural networks, medical imaging, smart cameras, image processing kernels, simulation of physical systems relevant to realistic visualization, as well as the solution of partial differential equations (PDEs) that arise in many cyber-physical systems such as automobile control and avionics.

Second, we target NVIDIA GPUs, which are \emph{vector-parallel programmable accelerators}.  Such components are now becoming de-facto standard in most embedded platforms and MPSoCs since they provide lightweight parallelism and energy/power efficiency.  We further argue that they will become ubiquitous for the following reasons.  Any device on the market today that has a screen (essentially, \emph{any device}, period) has to render images.  GPUs are natural platforms for this processing (for speed and efficiency).  So all systems will have an accelerator, by default.  If the system now needs any additional dense stencil computations, the natural target for performing it in the most speed/power/energy efficient manner is on the accelerator.

The third element of domain specificity is that we exploit a formalism called the \emph{polyhedral model} as the tool to map dense stencil computations to GPU accelerators.  Developed over the past thirty years~\cite{sanjay-fst-tcs, quinton-jvsp89, feautrier91, feautrier92a, feautrier92b}, it has matured into a powerful technology, now incorporated into \texttt{gcc}, \texttt{llvm} and in commercial compilers [Rstream, IBM].  Tools targeting GPUs are also available~\cite{grosser-etal-GPUhextile-CGO2014, chen_etal08_ChillTR}.

Thus, we formulate the domain-specific optimization problem: \emph{simultaneously optimize compilation and hardware/architectural parameters to compile stencil computations to GPUs}.

Previously, we presented~\cite{prajapati2017techreport} an approach to solve the above problem as follows:
\begin{enumerate}
\item \textbf{Develop Models}
	\begin{enumerate}
	\item \textbf{Time Model~\cite{prajapati2017simple}} We show that the elements of the domain specificity can be combined to develop simple, analytical (as well as accurate) models for the execution time of tiled stencil codes on GPUs and that these model can be used to solve for optimal tile size selection. Our model was able to predict tile sizes that achieve 30\% of theoretical machine peak on NVIDIA Maxwell GTX 980 and Titan~X.
	\item \textbf{Area Model~\cite{prajapati2017techreport}} We develop a simple, analytical model for the silicon area of programmable accelerator architectures, and calibrate it using the NVIDIA Maxwell class GPUs.  Our model proved to be accurate to within 2\% when validated.
	\item \textbf{Energy Model~\cite{prajapati-ipdpsw-energy}} We also developed energy models, as an explicit analytic function of a set of compiler and hardware parameters, that predict the energy consumption by analyzing the source code.  We used these energy models to obtain optimal solutions to tile size selection problem.
	\end{enumerate}
\item \textbf{Codesign~\cite{prajapati2017techreport}} We combine the proposed execution time model~\cite{prajapati2017simple} and the area model ~\cite{prajapati2017techreport} with a workload characterization of stencil codes to formulate a mathematical optimization problem that minimizes a common objective function of all the hardware and compiler parameters.  We propose a set of Pareto optimal designs that represent optimal combination of the parameters that would allow up to 126\% improvement in performance (GFlops/sec).
\end{enumerate}

Despite domain specificity, the problem remains difficult.  Even when done by hand for single target architecture and an application kernel, it is more art than science. Although smart designers and implementers have worked for many decades on such problems for the ``application/architecture du jour,'' each one was usually a point-solution~\cite{Pouchet-etalFPGA2013, Zuo-etal-ICCAD2013, Zuo-etal-CODES+ISSS2013}.  Designers invested blood, sweat and tears to find the best implementation, used it to solve their problem of interest, usually published a paper explaining the design, and moved on.  Their invested effort, particularly the trade-offs they made, and lessons they learned, are lost: future designers are left to reinvent wheels.

The high-level objective is to optimize stencil codes while tuning the hardware accelerator (GPUs) developing a complete ecosystem.  The goal is to \emph{\textbf{automatically and provably optimally, using time and/or energy as the objective function, map stencils to the hardware accelerators}}.  The idea is to obtain \emph{provably optimal} mappings through rigorous mathematical optimizations. \textbf{The proposed approach can have the following benefits.}
\begin{itemize}\itemsep 0mm
\item \emph{\textbf{Automation with Optimality}:} the most time/power/energy efficient implementations can be derived, reducing programmers' effort. \emph{Compilation tools} can be used to guide the optimal choice of transformations which will, in turn, \emph{optimize the performance} of the workloads such as deep learning, image rendering, cyber-physical systems, autonomous vehicle systems, etc.
\item \emph{\textbf{Future proofing:}} Porting applications to new GPU architectures will require less effort.  Instead of a redesign of each program, our methods can be used to develop new parallelization strategies and transformations, refine/redefine objective functions and constraints, and re-target the compiler.  This one-time effort can then be amortized over many application kernels.
\item \emph{\textbf{Codesign:}} By casting hardware/architectural parameters as \emph{variables} in the mathematical optimization framework, we can \emph{solve} for their optimal values.  This will enable us to systematically explore alternate GPU architectures and simultaneously tune compilation parameters. Such a codesign approach will help speed up the research work and the chip design process. The cost models can be used to quickly recognize the performance sinks and help \emph{identify the design flaws in its early stages saving billions of dollars}.
\end{itemize}
\paragraph{Generalization}
Future exascale high-performance computing (HPC) systems are expected to be increasingly heterogeneous, consisting of several multi-core CPUs and a large number of accelerators, special-purpose hardware that will increase the computing power of the system in a very energy-efficient way~\cite{exascale-dally2011power}.  Consequently, highly specialized coprocessors will become much more common in exascale than in the current HPC systems.  Such specialized, energy-efficient accelerators are also an important component in many diverse systems beyond HPC: gaming machines, general purpose workstations, tablets, phones and other media devices.  In order to attain exascale level performance, accelerators have to become even more energy-efficient, and experts anticipate that a large part of this must come through increased specialization~\cite{gameover}. 
%This paper presents a methodology for designing these accelerators.

%(HPC) \emph{System codesign} is one of the enabling technologies for exascale computing and beyond~\cite{b1282de6ab8a4f1a8a962cb40310bc7f}.  Currently, hardware and software optimization is done largely separately.  Hardware manufacturers design and produce a high-performance computing (HPC) system and deliver it to users, who then try to adapt their application codes to run on the new system.  But because of incompatibility between hardware and software parameters, often such codes are only able to run at a small fraction of the total performance the new hardware offers.  Hence, optimizing both the hardware and software parameters simultaneously during design is considered as one way to achieve better hardware usage efficiency and thereby enabling leadership-class HPC at a more manageable cost and energy usage.

Our approach can be used to solve the problem of \emph{System codesign} by applying proposed accelerator codesign techniques to all the classes of programs that optimize for all the parameters simultaneously.  We provide a proof of concept of our approach, which is a stepping stone towards solving the larger problem of \emph{transforming the GPUs into accelerators for HPC Systems.}
\section{Breaking Abstractions}
\label{sec:abstractions}
%%%%%%%%
%
\begin{figure}[ht]
\includegraphics[scale=0.4]{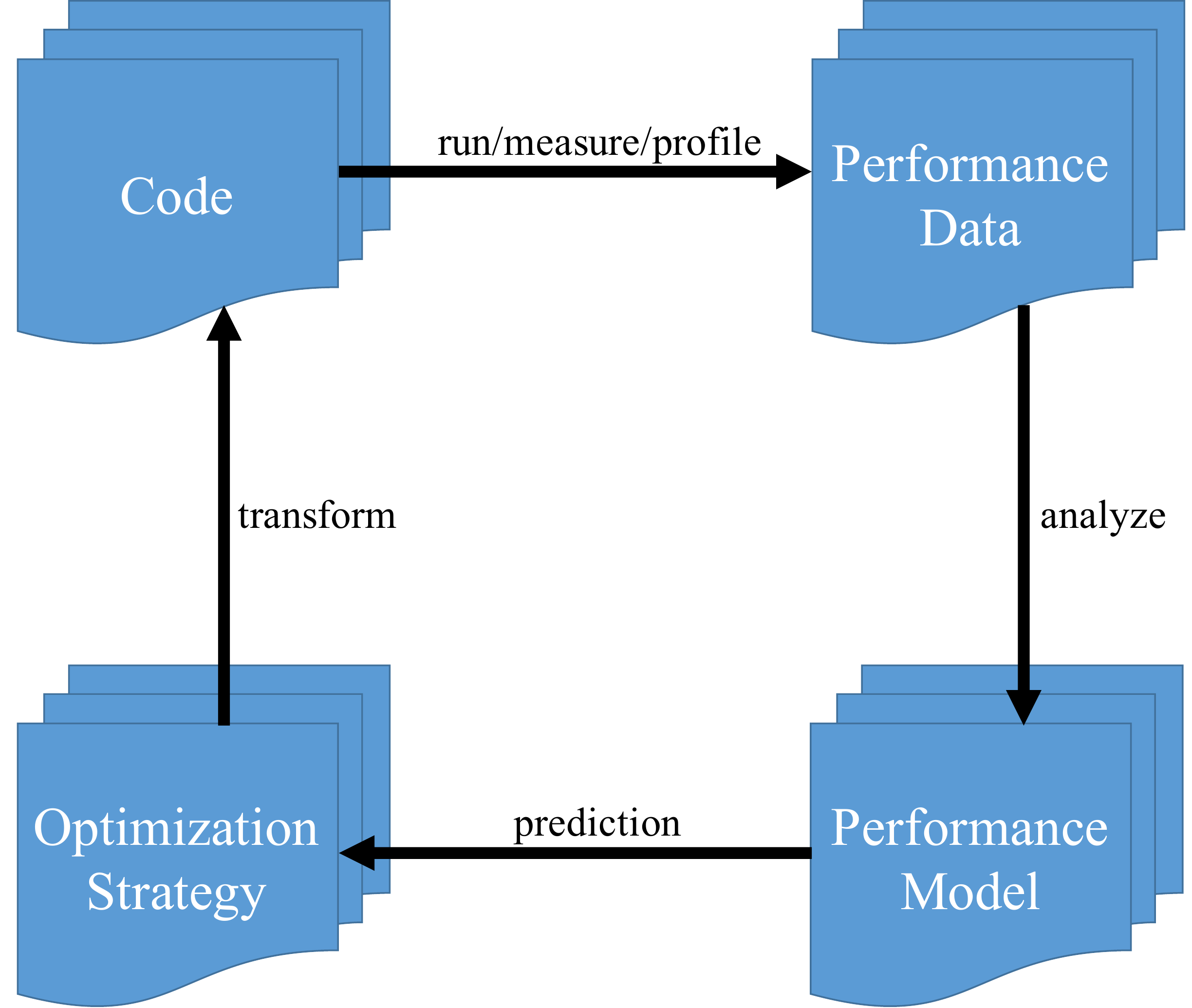}
\caption{HPC application performance improvement cycle. The layers represent different architectures.}
\label{fig:hpc-cycle}
\end{figure}
Since a long time the HPC developers and tool builders are using certain abstractions to improve the performance of applications.  Figure~\ref{fig:hpc-cycle} shows the performance improvement life cycle of an HPC application.  The layers represent a separate life cycle for every architecture (e.g. GPUs, CPUs, FPGAs).  A different Programming Language and System, each with its own Programming Environment \& Tools is used based on the underlying hardware architecture.  Therefore, there are different codes for the same application on different hardware.  Usually, the scientists who develop the algorithms/applications are completely isolated from the HPC performance tuning specialists.  For every application, a profiler is used to get performance data which is analyzed to derive a performance model.  The performance models are used to predict the performance and select an optimization strategy.  Optimal transformation strategy is applied to get code. This cycle repeats until the satisfactory performance is obtained. 
\begin{figure}[ht]
\includegraphics[scale=0.65]{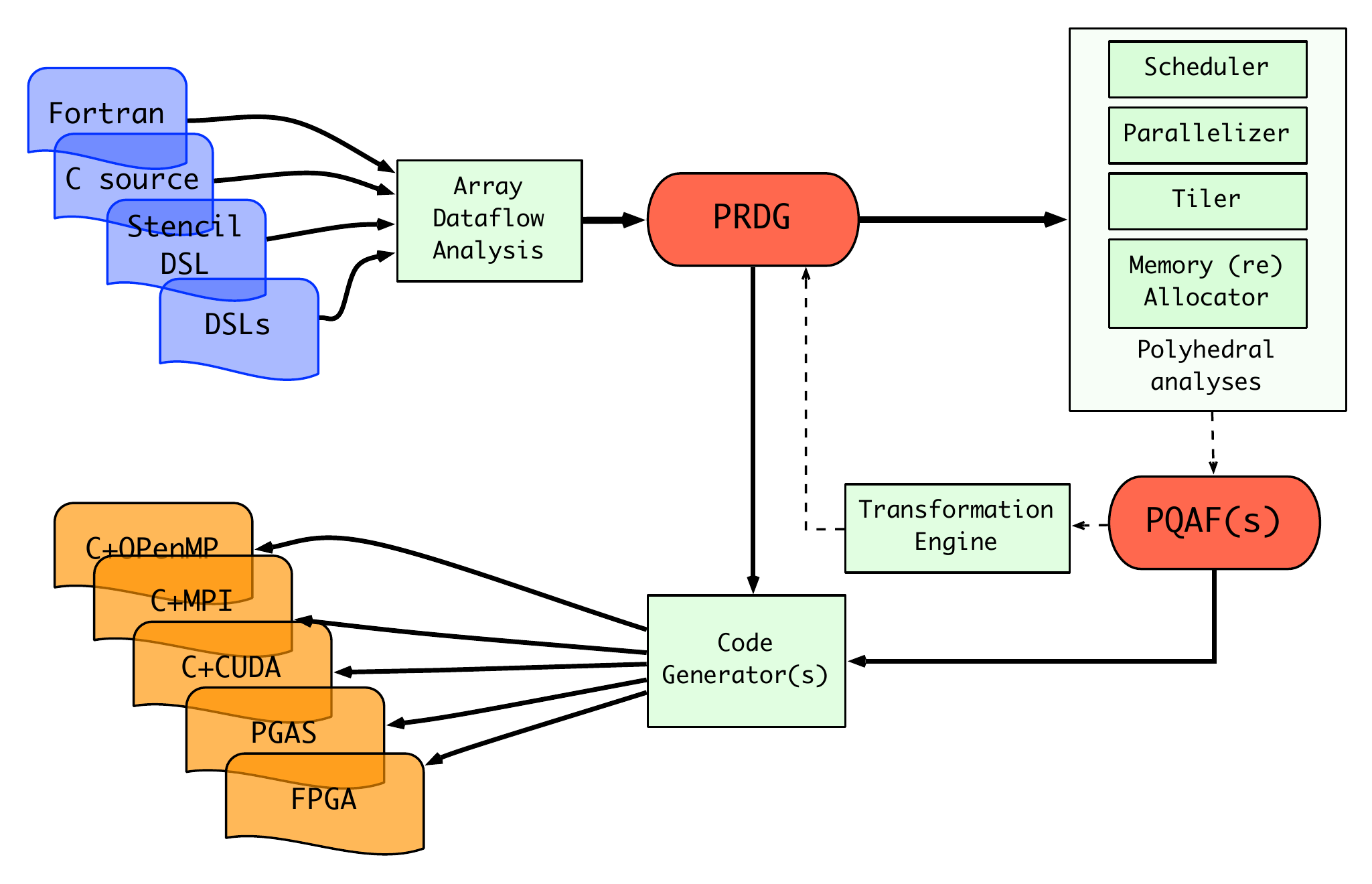}
\caption{Polyhedral Compilation}
\label{fig:polyhedral-compiler}
\end{figure}

Polyhedral compilers (see Figure~\ref{fig:polyhedral-compiler}) have a Polyhedral Reduced Dependence Graph (PRDG) as the intermediate representation of loops.   Polyhedral analysis (such as scheduling, parallelizing, memory allocation, etc.) is performed on this graph. Piecewise Quasi-Affine Functions (PQAFs) are mathematical functions that transform the PRDG using cost functions.  A transformed PRDG is used to generate codes.   Notice the similarities and differences between Figures~\ref{fig:hpc-cycle} and ~\ref{fig:polyhedral-compiler} and their performance cycles.  The optimization strategies in Figure~\ref{fig:hpc-cycle} are represent the PQAFs.  The performance models are subsumed in the polyhedral analysis phase.  Our work focuses on using performance models for polyhedral analysis, in turn, breaking the cycle and reducing the time to find optimal solutions.
\paragraph{Novelty}
\begin{enumerate}
\item The main novelty of our work comes out as a consequence of some of the exascale challenges~\cite{gameover}.  For exascale system design, various architectures, programs and transformation strategies are to be explored simultaneously in order to find the optimal.  We add performance models to this design space and provide \emph{\textbf{a unified view of the optimization space}}.  Figure~\ref{fig:view} shows this view (more details in Section~\ref{sec:prob}). 
\begin{figure}[ht]
\includegraphics[scale=0.33]{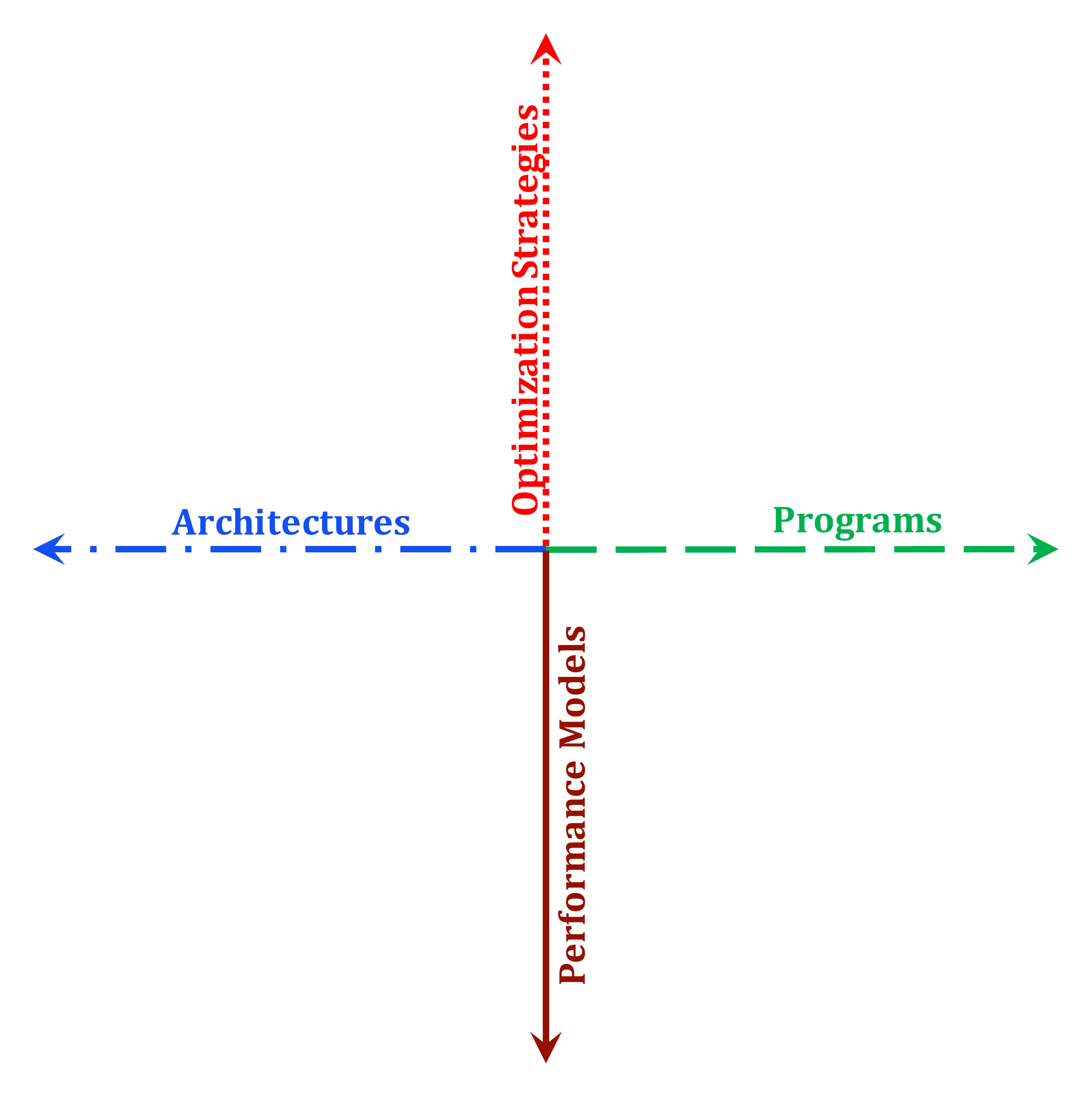}
\caption{Unified View of the Design Space.}
\label{fig:view}
\end{figure}
\item The above design space, however, is very large, has too many parameters and is too complicated to develop precise models.  Therefore, we explore \emph{\textbf{domain specificity and identify regions}} where optimization across multiple axes become possible.  
%
%two specific elements of PARRIC are especially novel and will merit special validation.  First of all, PARRIC further raises the level of abstraction to allow the programmer's specification to be highly compact, as \emph{\textbf{polyhedral equations}}.  We claim that this greatly decreases the programming effort of the domain scientist, 
%and we will validate this claim quantitatively with three experiments.  First we will compare the actual size of the starting specification and quantify how compact sloppy equations are.  Next we will measure the reduction in development time that can be achieved with these tools.  And finally, we will deploy and derive new algorithms for three different RRI tools: the nupack family of algorithms, Sankoff's algorithm, and extension of \emph{\textbf{piRNA}} to multiple sequences.
%
%Indeed, PI Rajopadhye has spent his entire research career spanning over thirty years arguing that polyhedral equations are the appropriate abstractions to enable target agnostic tools for systematic parallelization of such programs to a variety of platforms, ranging from multicore processors and distributed memory machines, to dedicated hardware like ASICs and FPGAs to general purpose where the micro-architecture is also synthesized by the ``compiler.''
\end{enumerate}
%%%%%%%%%%%%%%%%%%%%%%%%%
We show how the analytical cost models can be used to optimize the performance of domain specific programs using transformation strategies for a given architecture in Chapter~\ref{chap:approach}. The rest of this document is organized as follows:  Chapter~\ref{chap:approach} explains our proposed approach.  Chapters~\ref{chap:costmodels},~\ref{chap:autotuning}, and~\ref{chap:codesign} discuss the work that has been accomplished.  Bottleneck Analysis is explained in details in Chapter~\ref{chap:bottleneckology}.  Chapter~\ref{chap:related} discusses the relevant literature.  Finally, Chapter~\ref{chap:conclusions} concludes the work.

% + Motivation - What is the problem?
% + Significance - why is it important? 

% + Abstractions - old and new
% + Challenges - Why is the problem difficult?
% + Objectives of the project.
% + Timeline

\chapter{The Landscape and Navigation}
\label{chap:approach}
%%%%%%%%%%%%%%%%%%%%%%%%%%%%%%%%%%%%%%%%%%%%%%%%%%%%%%%%%%%%%%%%
%
\section{Domain Specificity}

As we move to address the challenges of exascale computing, one approach that has shown promise is \emph{domain specificity:} the adaptation of application, compilation, parallelization, and optimization strategies to narrower classes of domains.  An important representative of such a domain is called \emph{Stencil Computations}, and includes a class of typically compute bound parts of many applications such as partial differential equation (PDE) solvers, numerical simulations in domains like oceanography, aerospace, climate and weather modeling, computational physics, materials modeling, simulations of fluids, and signal and image-processing algorithms.  One of the thirteen Berkeley dwarfs/motifs~\cite{asanovic-etal-berkeley-view2006}, is ``structured mesh computations,'' which are nothing but stencils.  Many \emph{dynamic programming} algorithms also exhibit a similar dependence pattern. The importance of stencils has been noted by a number of researchers, indicated by the recent surge of research projects and publications on this topic, ranging from optimization methods for implementing such computations on a range of target architectures, to Domain Specific Languages (DSLs) and compilation systems for stencils~\cite{datta-etal-sc08, dursun-etal-europar09, dursun-etal-pdpta09, kamil-etal-mspc06, kamil-etal-msp05, KBBRRS-stencil-pldi07, liu-etal-ipdps09, micikevicius-gpgpu09, nitsure-ms06, strzodka-etal-ics10, strzodka-etal-ppopp11-poster, Tang2011Pochoir, shaheen-ipdps12}.  Workshops and conferences devoted exclusively to stencil acceleration have recently emerged.

\nocite{epperson-num-methods07, bleck-etal-JPO92, griffies-etal-ocean2000,
  john-options-text-06, mei-etal-LBM2000, nakano-etal-CPC94,
  chowdhury-etal-TCBB10, sanjay-gpugems10}

A second aspect of domain specificity is reflected in the emergence of specialized architectures, called \emph{accelerators}, for executing compute intensive parts of many computations.  They include GPGPU, general purpose computing on graphics processing units (GPUs), and other co-processors (Intel Xeon Phi, Knight's Landing, etc.). Initially they were ``special purpose,'' limited to highly optimized image rendering libraries occurring in graphics processing.  Later, researchers realized that these processors could be used for more general computations, and, eventually, the emergence of tools like CUDA and OpenCL enabled general purpose parallel programming on these platforms.

Exploiting the specificity of the applications and the specificity of target architectures leads to domain-specific tools to map very high level program specifications to highly tuned and optimized implementations on the target architecture.  Many such tools exist, both academic research prototypes and productions systems.
%They include blah, blah, and ppcg.

As indicated earlier, our domain specificity comes in multiple flavors. First, we investigate only stencil computations.  They belong to a class of programs called \emph{uniform dependence computations}, which are themselves a proper subset of ``affine loop programs.''  Such programs can be analyzed and parallelized using a powerful methodology called the polyhedral model~\cite{sanjay-fst-tcs,quinton-jvsp89,quinton-sanjay-tf,feautrier91,feautrier92a,feautrier92b,Darte-book,uday-pldi08}, and many tools are widely available, e.g., PPCG, developed by the group at ENS, Paris~\cite{verdoolaege2013polyhedral}.  Second, we tackle a specific target platform, namely a single GPU accelerator, and PPCG includes a module that targets GPUs and incorporates a sophisticated code generator developed by Grosser et al.~\cite{grosser-etal-GPUhextile-CGO2014} that employs a state-of-the-art tiling strategy called \emph{hybrid hexagonal classic tiling}.  An open source compiler, implementing this strategy is also available, henceforth called the HHC compiler.

%\paragraph{Why stencils?}
%Stencil computations are ubiquitous and important as recognized by many authors~\cite{datta-etal-sc08, dursun-etal-europar09, dursun-etal-pdpta09, kamil-etal-mspc06, kamil-etal-msp05, KBBRRS-stencil-pldi07, liu-etal-ipdps09, micikevicius-gpgpu09, nitsure-ms06, strzodka-etal-ics10, strzodka-etal-ppopp11-poster, shaheen-ipdps12, pochoir-spaa11, frigo-strumpen-ics05, frigo-strumpen-tocs09}.  Domain specific stencil languages and compilers have been extensively researched, for over two decades~\cite{SciNapse97, cactus00, Bassetti98, brickner-etal-cpc93, bromley-etal-pldi91, pochoir-spaa11, Decker98}. %, but there is now a renewed interest from diverse application domains from .

\subsection{Comparison with Polyhedral Methods}
The landscape described (in Figure~\ref{fig:view}) allows us to place our work in context.  Although our methods are for domain specific purposes, an extreme situation with CPUs as the architecture, and the set of polyhedral programs allows us to compare with conventional compilation.

The optimization problem a compiler ``solves'' is: \emph{pick transformation parameters so as to optimize the program property of interest, typically execution time.}  Since it has a single (or a small handful of) predetermined strategies, it is a limited kind of mathematical optimization problem.  The objective function is a surrogate for execution time.
%, and the solution is often a local, rather than global optimum.

Now consider PLuTO~\cite{uday-pldi08}, a state-of-the art polyhedral compiler based on a mathematical representation of both programs and transformations. By considering only polyhedral programs and transformations, the optimization problems are rigorous.  By default, PLuTO targets multicore CPUs, and uses a transformation strategy that combines one level of tiling, loop fusion and (loop/wavefront) parallelization of tiles.  It solves a mathematical optimization problem where the schedule parameters (coefficients of tiling and schedule hyperplanes) are the unknown variables, and the cost function is the number of linearly independent tiling hyperplanes, combined with a quantitative measure of the length of the inter-tile dependences.  This is again, a surrogate for the total execution time, and leads to solutions that while reasonable, are not \emph{provably} optimal.  Moreover, parameters like tile sizes, vectorization and inter-tile schedule are chosen using simple heuristics, and are not part of the optimization.

\subsection{Limitations of current domain-specific compilation}
\label{sec:probdomain}

Consider how polyhedral compilation has recently evolved.  Bondhugula et al.~\cite{uday-pact2014} proposed an extension of PLuTO for periodic stencil computations, and Bandishti et al.~\cite{bandishti12} developed another extension to allow concurrent starts.  Since the objective functions in these strategies are all surrogates for the execution time, there is no way to compare across the strategies.  Authors leave the choice of strategy to the user, via compiler flags.  Recently it was shown (in ~\cite{yun-dissertation2016}, both quantitatively and empirically) that while concurrent start may be faster for iteration spaces with a certain aspect ratio of the program size parameters, the best performance for the same program with different aspect ratio is provided by the basic PLuTO algorithm.

As another example, Grosser et al.~\cite{grosser-etal-GPUhextile-CGO2014} proposed a novel combination of hexagonal and classic tiling for stencil programs on GPUs.  They demonstrated---only empirically---performance gains compared to previous strategies, but did not quantitatively explore cases where HHC was better.

As a consequence, polyhedral compilation remains difficult.  Every time a new strategy is developed, the authors publish a paper, and empirically show that their results are better than previous ones.  They usually do not provide a quantitative, analytical comparison, thereby preventing a better, collective understanding of how to solve the bigger, global problem.  Our intention is not to criticize the field: the problems today are difficult enough that significant effort is needed for even developing such ``point solutions.'' Our approach is a step towards addressing these limitations.

% Local Variables: ***
% TeX-master: "../stencil-time-model-2d.tex" ***
% fill-column: 78 ***
% End: ***

%
% CHACK - design space + energy
% Lakshmi's thesis
% Research Objectives - THESIS STATEMENT
\section{Design Landscape}
\label{sec:prob}

\begin{figure}[ht]
  \centering
  \includegraphics[width=7.7cm]{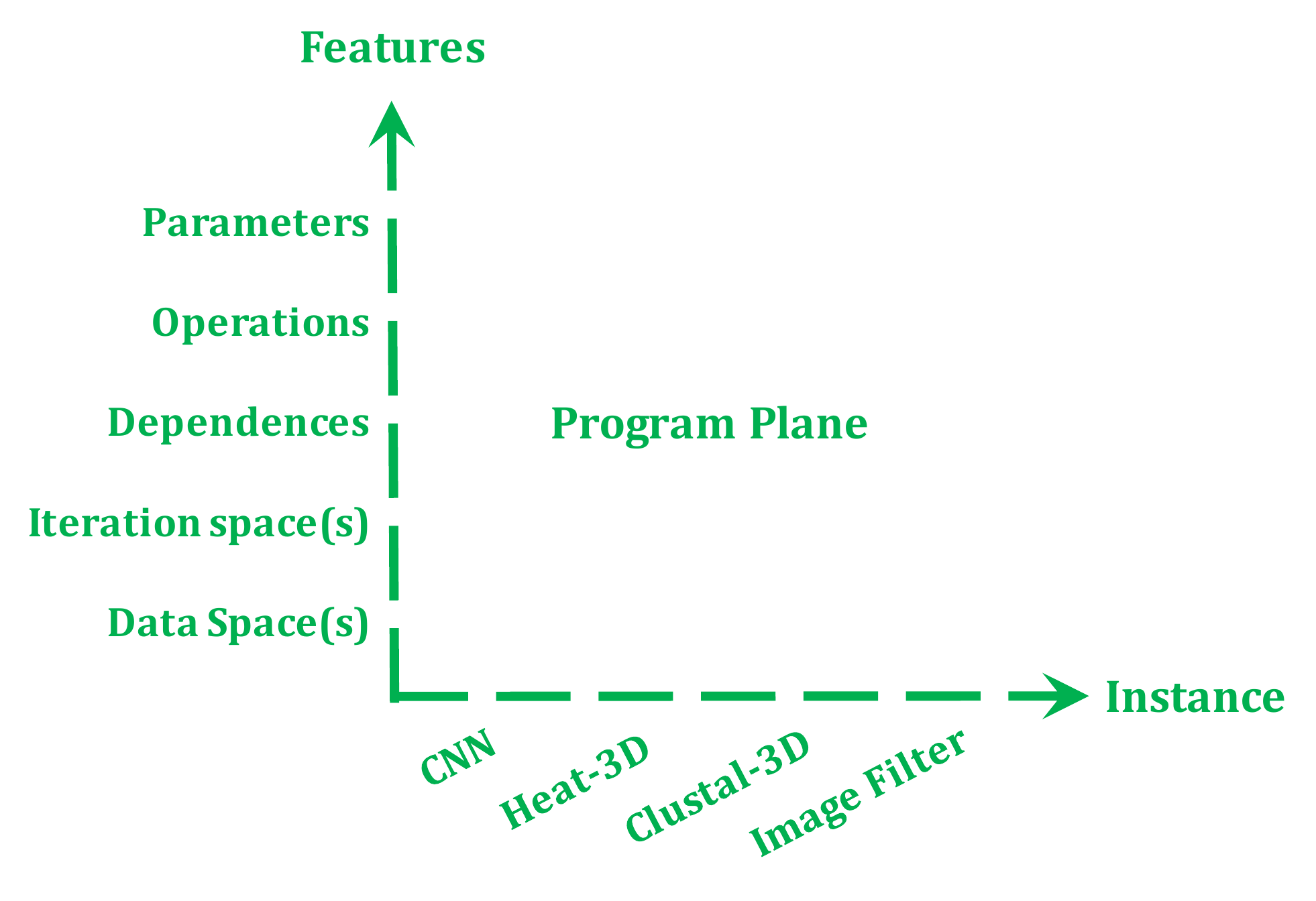} ~~~~
  \includegraphics[width=7cm]{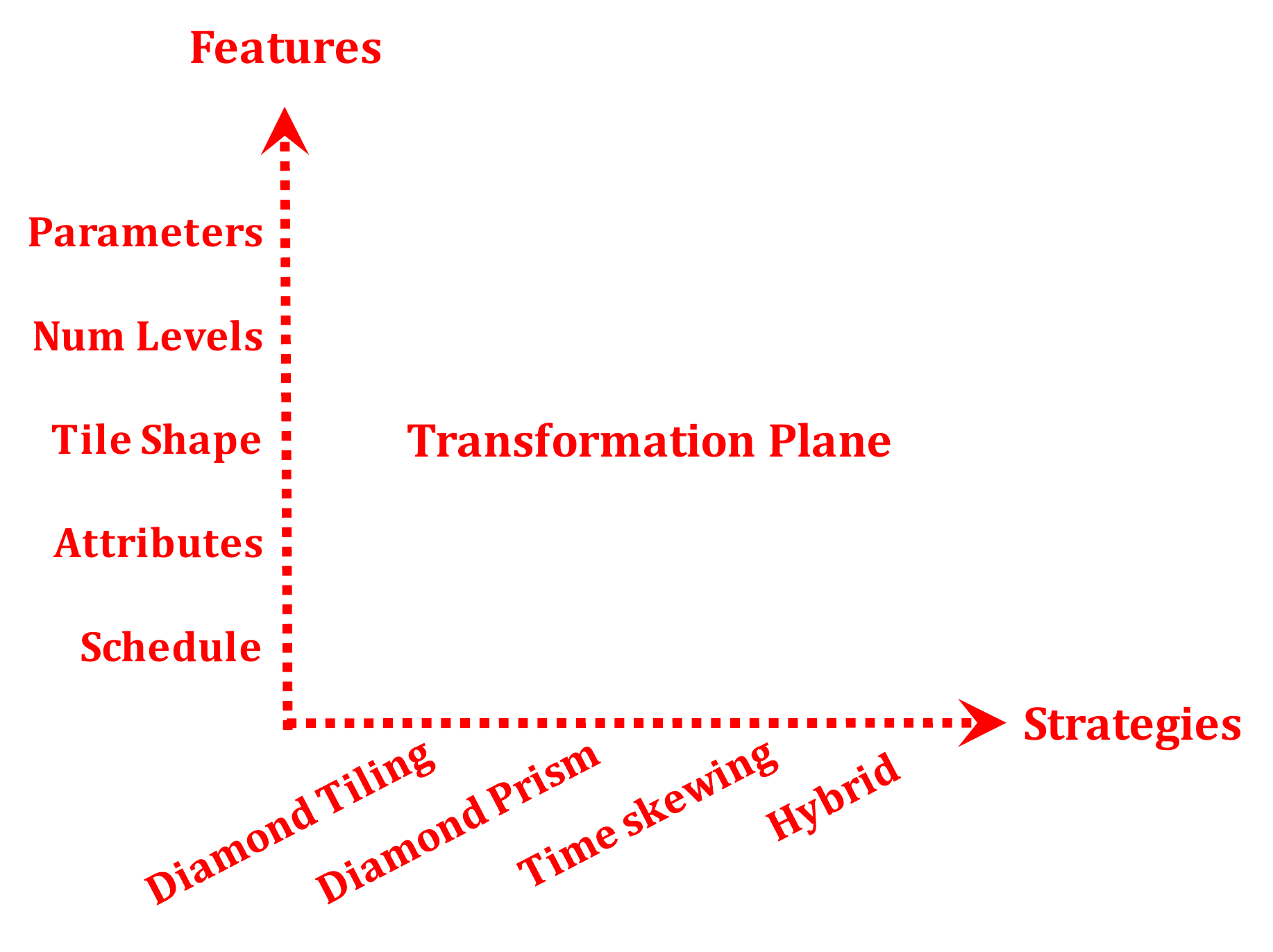}

  \centerline{\mbox{}\hfill (a) Programs \hfill\hfill (b) Transformations \hfill
    \mbox{}}
  \includegraphics[width=7.7cm]{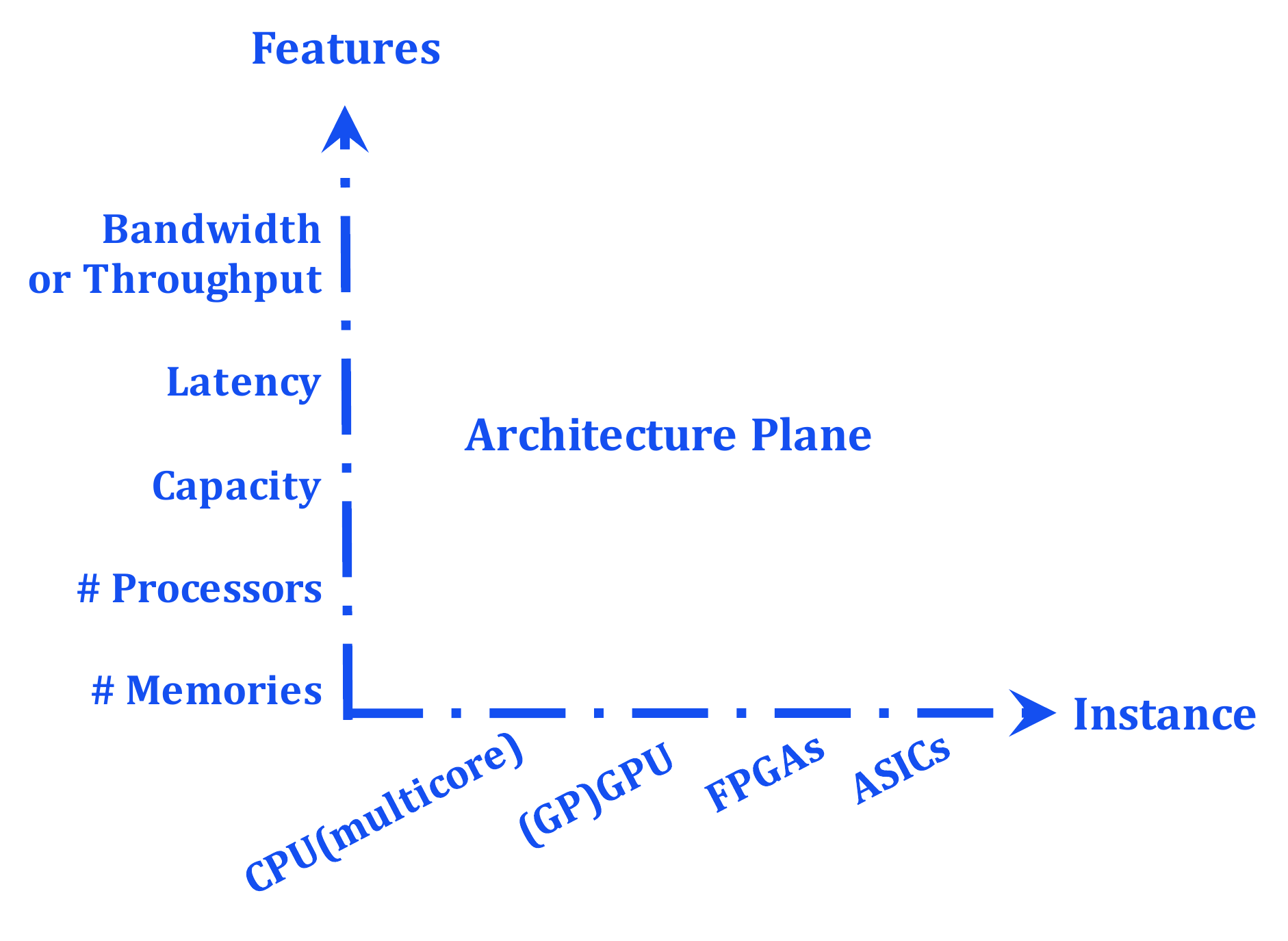} ~~~~
  \includegraphics[width=7cm]{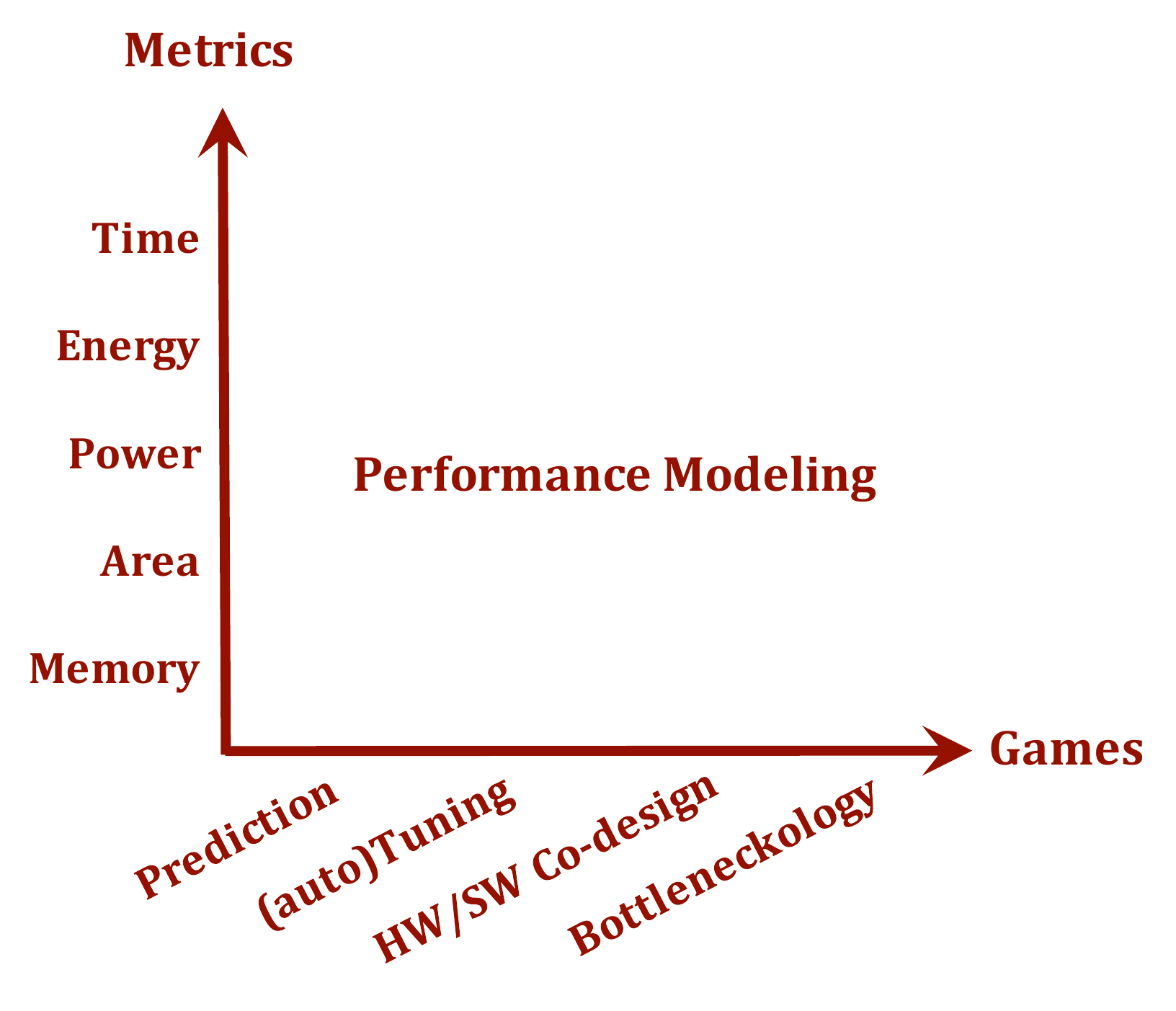}

  \centerline{\mbox{}\hfill (c) Architectures \hfill\hfill (d) Performance Models \hfill
    \mbox{}}
  \caption{\small{The Domain-Specific Design Landscape.  We tackle a narrow domain of programs (a) each one described by a small number of features. Similarly the mappings/transformations (b) we can apply are drawn form a small set, each one parameterized by a set of features.  Architectures (c) are also drawn from a similarly small set, and are parameterized by their features.  In each plane, the features may be hierarchical. The performance models (d) and the games that we can play. }}
  \label{fig:landscape}
\end{figure}

To place our work into context, to precisely formulate the problems we address, and to describe the approach we take to solve them,  we show the design landscape of domain-specific optimization problems.  It has six dimensions, organized into three planes (Fig.~\ref{fig:landscape} (a), (b) and (c)).  Each plane has two axes: \emph{instances}, and \emph{features}.  The feature axis may be hierarchical, and parameterized.

The \emph{\textbf{program plane}} (Fig.~\ref{fig:landscape}(a)) consists of instances of \emph{\textbf{dense stencil computations}}, such as \emph{\textbf{Convolutional Neural Net}} (a machine learning kernel), \textbf{\emph{Heat-3D}} (a stencil computation from computational science),  \emph{\textbf{Clustal-3D}} (a dynamic programming kernel from bioinformatics).  Because of domain specificity, each program is \emph{compactly} described with a small set of \emph{features}, such as: (i) a set of \emph{iteration spaces} (ii) a set of \emph{data spaces}, (iii) a set of \emph{dependences}, (iv) a set of \emph{computational operators} (e.g., loop bodies), and 
%optionally, \footnote{Schedules are optional, since they may be implicit in purely declarative programs.  Any (partial/total) order that respects the dependences is legal, and it is the compiler's job to provide/change the schedule.} (v) a set of \emph{schedules}.  A program instance may have 
(iv) one or more \emph{size parameters}.

The \emph{\textbf{transformation plane}} (Fig.~\ref{fig:landscape}(b)) defines the space of compiler transformations that can be applied to the program.  Domain specificity again allows us to consider only a few instances, e.g., time skewing~\cite{wonnacott-ijpp2002, sanjay-tpds03}, diamond tiling~\cite{bandishti12}, diamond prisms~\cite{strzodka-etal-ics10, StShPa_11CATS}, or hybrid hexagonal-classic (HHC) tiling~\cite{grosser-etal-ppl2014, grosser-etal-topla2015, grosser-etal-GPUhextile-CGO2014}.  Transformation strategies are (potentially) hierarchical, and each level of the hierarchy represents a \textbf{\emph{partitioning}}.\footnote{\emph{Partitioning} denotes a generalization of a crucial transformation called \emph{\textbf{tiling}} to also include \emph{\textbf{multiple passes}}.}
They are also specified by a set of \emph{features}, each of which is a mathematical function: (i) \emph{tile shape}, specified by the so called ``tiling hyperplanes,'' (ii) \emph{tile schedule}, (iii) \emph{processor mapping} specifying which (virtual/physical) processor in the hardware hierarchy will execute a tile, and (iv) \emph{memory allocation} specifying where its inputs and outputs are stored.  Note that the schedule usually also has components to specify when tile inputs are read and when tile outputs are written.  The transformation plane features are also parameterized: mapping function coefficients, tile sizes, etc., are viewed as parameters.

The \emph{\textbf{architecture plane}}  (Fig.~\ref{fig:landscape}(c)) captures the accelerator hardware.  Examples of architecture instances are ASICs, FPGAs/Reconfigurable Logic, and Instruction Set Architectures (such as GPGPUs).\footnote{ In our taxonomy, an accelerator that is programmable in the conventional, von Neumann sense is an ISA, and accelerators where the ``programming'' is at the circuit level are FPGAs or ASICs.} 
%The features in the architecture plane represent the hardware hierarchy
%~\cite{alpern-etal-UMH-model1994, alpern-etal-PMH-model1993}.  At each level are parameters of the hardware such as 
Hardware features include (i) the number of processors at that granularity, (ii) the memories, and (iii) the interconnects.  Parameters of each such feature specify the performance of that feature: e.g., speed in terms of throughput and latency, capacity, etc. %as well as the power/energy cost including both dynamic (active) and static (leakage) energy.
Also included are specific access/scheduling constraints such as the need for gang scheduling in warps, the constraints on coalescing and bank conflicts, etc.
\begin{figure}[ht]
  \centering
  \includegraphics[width=14cm]{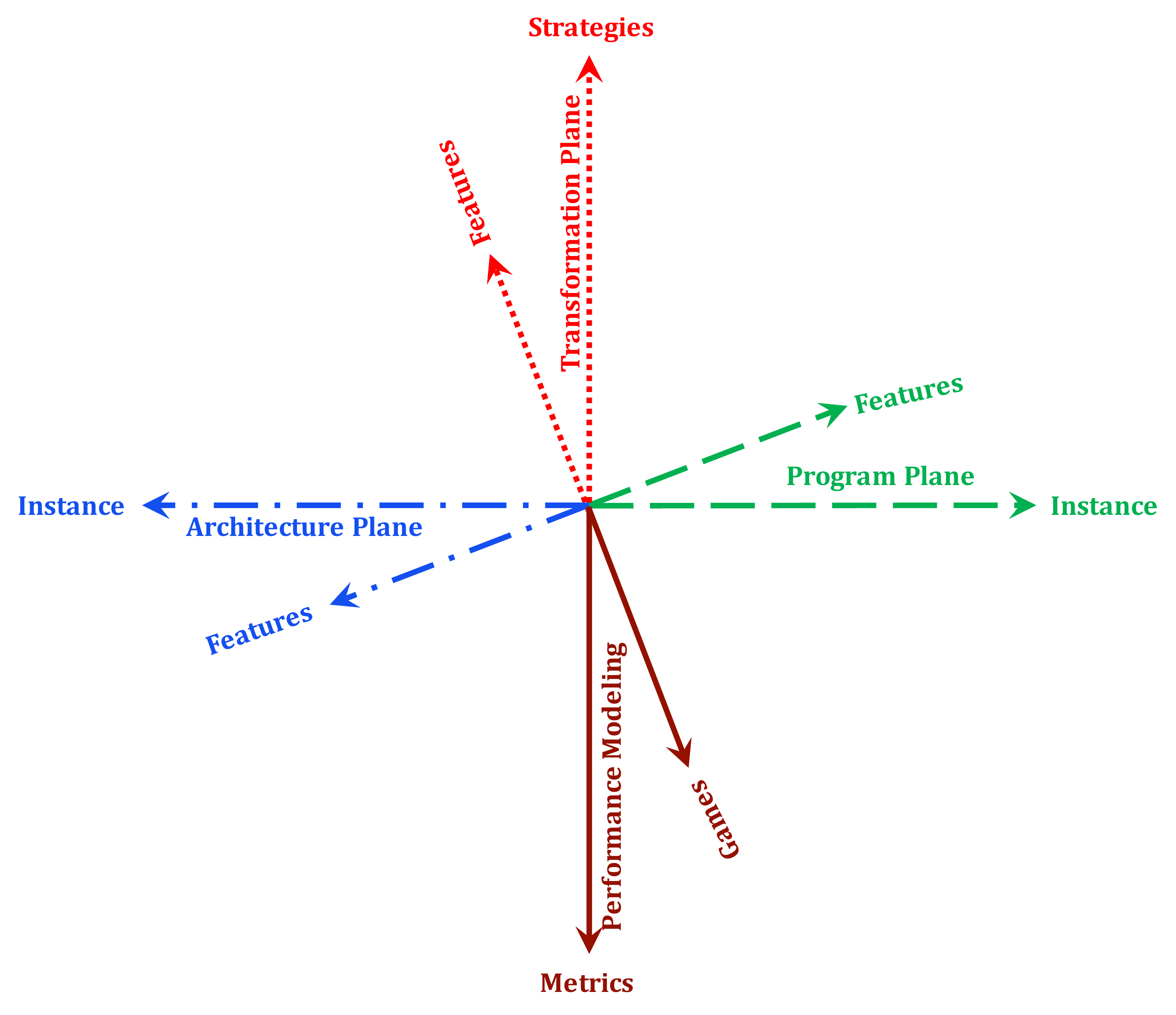} ~~~~~
   \caption{\small{The main novelty of our work! The Domain-Specific Design Landscape that we must navigate is thus the cross-product of the four planes in Fig.~\ref{fig:landscape}.}}
  \label{fig:designspace}
\end{figure}

We combine this six-dimensional design space with the \emph{\textbf{performance modeling plane}} (Fig.~\ref{fig:landscape}(d)) which gives us a high level picture of various performance metrics and the games that we can play with these performance models.  \emph{\textbf{This unified view of the landscape is the main novelty of our work}}. Fig.~\ref{fig:designspace} shows the eight dimensional navigation space to be explored.  A model can be used for performance \emph{\textbf{prediction}} as well as performance improvement like \emph{\textbf{(auto)tuning}}, \emph{\textbf{bottleneckology}} (bottleneck analysis), etc.  Multiple models can contribute to \emph{\textbf{multi-criteria optimization}} (eg. simultaneously optimize for both, time and energy) as well as \emph{\textbf{HW/SW Co-design}}.

% In general, each new strategy is proposed as a singular result, and the
% community does not have a collective comparative understanding.

% One of the goals of our project is that such code generators would simply be
% instances of a single ``meta-level'' code generator.

% Local Variables: ***
% TeX-master: "CHACK.tex" ***
% fill-column: 78 ***
% End: ***

%
%
%
\section{Approach}
We now describe our overall approach and how it can lead to the benefits mentioned earlier (i.e., \emph{automatic optimal mappings}, \emph{future proofing} and \emph{codesign}).  First, we develop analytical models for execution time and energy for a given program and a transformation strategy on a fixed architecture.  We also develop silicon area models for GPU architectures and show its use in chip area prediction.  Second, we show how these models can be used for performance optimization.  And finally, we show how to formulate mathematical optimization problems using such cost models to solve the problem of software-hardware codesign.  We show our initial results to justify our claims, and identify remaining challenges in later chapters.
\subsection{Models and Validation}
\subsubsection{Execution Time Model and Prediction of GPGPU Stencils}
%Stencil computations are an important class of compute and data intensive programs that occur widely in scientific and engineering applications.  A number of tools use sophisticated tiling, parallelization, and memory mapping strategies, and generate code that relies on vendor-supplied compilers.  This code has a number of parameters, such as problem sizes and tile sizes.  
We develop an execution time model that predicts execution time of transformed stencil codes.  Our model is a simple set of analytical functions that predict the execution time of the generated code.  It is deliberately optimistic, since we are targeting modeling and parameter selections yielding highly optimized codes.  We experimentally validate the model on a number of 2D and 3D stencil codes, and show that the root mean square error in the execution time is less than 10\% for the subset of the codes that achieve performance within 20\% of the best.
\subsubsection{Energy Model and Prediction of GPGPU Stencils}
Like the analytical execution time model, we develop a methodology for modeling the energy efficiency of tiled nested-loop codes running on a graphics processing unit (GPU) and use it for prediction of energy consumption.  We assume that a highly optimized and parameterized version of a tiled nested-loop code---either written by an expert programmer or automatically produced by a polyhedral compilation tool---is given to us as an input.  We then model the energy consumption as an analytical function of a set of parameters characterizing the software and the GPU hardware. Most previous attempts at GPU energy modeling were based on low-level machine models that were then used to model whole programs through simulations, or were analytical models that required low level details.  In contrast, our approach develops analytical models based on (i) machine and architecture parameters, (ii) program size parameters as found in the polyhedral model, and (iii) tiling parameters.  Our model therefore allows prediction of the energy consumption with respect to a set of parameters of interest.  We illustrate the framework on three nested-loop codes: Smith-Waterman, and one-dimensional and two-dimensional Jacobi stencils, and analyze the accuracy of the resulting models.  With optimal choice of model parameters the RMS error is less than 4\%. Two factors allow us to attain this high accuracy.  The first is domain-specificity: we focus only on tilable nested-loop codes. The second is that we decouple the energy model from a model of the execution time, a known hard problem.
\subsubsection{Area Model and Chip Area Prediction of GPUs}
We also develop an analytic model for the total silicon area of a GPU accelerator.  We faced some difficulties in deriving an acceptable analytical model, as silicon data had to be reverse engineered from extremely limited public domain resources.  As a general observation, within each GPU family, there is little diversity in the parameter configurations.  For the Maxwell family of GPUs, the GTX980 and Titan~X chips were chosen as two sufficiently distinct points to calibrate our analytical models.  The calibration itself was performed by evaluating die photomicrographs, publicly available information about the nVidia GTX 980 (Maxwell series) GPU, and other generally accepted memory architecture models.  The model validation was done by comparing the predictions with known data on the Maxwell series Titan~X GPU. We found the model prediction to be accurate to within, 2\%, though this number is not significant\footnote{Although a many configurations of any family of GPUs are spaced out, they come from binning only a small number of distinct dies.  We ended up calibrating our model on one die and validating it on only another one.}.

Next, we develop mathematical objective functions to illustrate the use of these models in performance optimization and later we will show the same for software-hardware codesign.
\subsection{Compilation and its optimization subspaces}
\label{sec:optimization}

To address the limitations of current domain-specific compilation noted in Section~\ref{sec:probdomain}, we now describe our approach to systematically exploring well defined regions of the design landscape using exact (not surrogate) objective functions.

\begin{enumerate}
\item \textbf{(Auto) Tuning:} 
Let us consider a three-dimensional subspace in our landscape, of \emph{instances} of programs $\times$ transformations/strategies $\times$ \emph{execution time}.  At each point in this three-dimensional space, we may define a mathematical optimization problem which has an \emph{objective function}, and a \emph{feasible space}.  Both involve analytical functions of parameters from all three feature dimensions, defined by vectors: $\vec{P}$ for a program, $\vec{S}$ for a strategy, and $\vec{A}$ for an architecture.  The objective function for execution time is $M_{T}(\vec{P}, \vec{S}, \vec{A})$.
%, and energy, $E(\vec{P}, \vec{S}, \vec{A})$.  
Similarly, we have constraints defining the feasible space of the optimization problem, ${\cal F}_T(\vec{P}, \vec{S}, \vec{A})$.
The mathematical optimization problems can be formulated for various performance metrics as follows:
\begin{enumerate}\itemsep 0mm
\item For execution time, minimize $M_T(\vec{P}, \vec{S}, \vec{A})$ subject to ${\cal F}_T(\vec{P}, \vec{S}, \vec{A})$, 
\item For energy, minimize $M_E(\vec{P}, \vec{S}, \vec{A})$ subject to ${\cal F}_E(\vec{P}, \vec{S}, \vec{A})$,
\item For power, minimize $M_P(\vec{P}, \vec{S}, \vec{A})$ subject to ${\cal F}_P(\vec{P}, \vec{S}, \vec{A})$, etc.
\end{enumerate}
Each problem instance has an objective function that represents (is not just a surrogate for) the metric which we seek to optimize: execution time($M_T$), power($M_P$), energy($M_E$), etc.  It is a function of all the parameters of this three-dimensional point.  Other parameters, e.g., the number of processors, the memory capacity, etc., may define a feasible space where this function is valid.

Our approach is based on the hypothesis that domain-specificity of both the programs and the architecture allows us to develop such functions.  Note that the objective function cannot be a surrogate, it must be the actual cost metric of interest.  Under this hypothesis, our entire strategy can be summarized as \emph{collective solution of multiple optimization problems with common objective function(s)}.  We will discuss two such common objective functions, ${\cal M}_T$ for execution time and ${\cal M}_E$ for energy (expressed as GOPs/sec or GOPs/joule) in details in the following chapters.

\item \textbf{(Auto) Super-Tuning:} 
%\paragraph{Optimizing across strategies:}
The next step will be to extend the optimization across multiple strategies, say $S_1$ and $S_2$.  Given two separate optimizations formulated as follows:
\begin{enumerate}\itemsep 0mm
\item Minimize $M_{T_1}(\vec{P}, \vec{S_1}, \vec{A})$ subject to ${\cal F}_{T_1}(\vec{P}, \vec{S_1}, \vec{A})$, and
\item Minimize $M_{T_2}(\vec{P}, \vec{S_2}, \vec{A})$ subject to ${\cal F}_{T_2}(\vec{P}, \vec{S_2}, \vec{A})$
\end{enumerate}
We can formulate the problem of optimizing across strategies in two ways: (i) Take the minimum of the two optimizations $min($minimize $M_{T_1}(\vec{P}, \vec{S_1}, \vec{A}),$ minimize $M_{T_2}(\vec{P}, \vec{S_2}, \vec{A}))$, or (ii) solve separate optimization problems, depending on the intersections and differences of the feasible spaces of each one.
\begin{enumerate}\itemsep 0mm
\item Minimize $\min(M_{T_1}(\vec{P}, \vec{S_1}, \vec{A}), M_{T_2}(\vec{P}, \vec{S_2}, \vec{A}))$, subject to ${\cal F}_{T_1}(\vec{P}, \vec{S_1}, \vec{A}) \cap {\cal F}_{T_2}(\vec{P}, \vec{S_2}, \vec{A})$,
\item Minimize $M_{T_1}(\vec{P}, \vec{S_1}, \vec{A})$, subject to ${\cal F}_{T_1}(\vec{P}, \vec{S_1}, \vec{A}) \cap \sim {\cal F}_{T_2}(\vec{P}, \vec{S_2}, \vec{A})$, and
\item Minimize $M_{T_2}(\vec{P}, \vec{S_2}, \vec{A})$, subject to $\sim {\cal F}_{T_1}(\vec{P}, \vec{S_1}, \vec{A}) \cap {\cal F}_{T_2}(\vec{P}, \vec{S_2}, \vec{A})$
\end{enumerate}
This can be extended to a set of strategies, ${\cal S} =\{ S_i\}$.  Although the second option is not very scalable---the number of sub-problems grows exponentially with the number of strategies---it is reasonable for a small number of strategies, e.g., it would let us automatically choose between time skewing, diamond tiling, diamond prisms, and HHC.  %Later, we will also discuss techniques to prune this space in a scalable manner in [GIVE CHAPTER].

\item \textbf{Multi-Metric (Auto) Tuning:} 
The above optimizations account for only one performance metric, which leads to a single objective function for the optimization.  One might want to optimize for more than one metric.  Let us consider a multi-metric optimization such as the \emph{energy-delay product}.  The optimization problem can be formulated as
\[ \mathrm{Minimize~~} (M_{T}(\vec{P}, \vec{S}, \vec{A}) \ast M_{E}(\vec{P}, \vec{S}, \vec{A})), \mathrm{~~subject~to~~} {\cal F}_{T}(\vec{P}, \vec{S}, \vec{A}) \cap {\cal F}_{E}(\vec{P}, \vec{S}, \vec{A}) \]
Note, the feasible space consists of the intersection of the feasible space of time and energy.  The program parameters (eg. problem sizes) and the features (eg. tile sizes)  of the selected strategy (eg. Diamond tiling) are the parameters to the multi-metric objective function.

\item \textbf{Multi-Metric (Auto) Super-Tuning:} 
The above multi-metric objective function can be extended to multiple strategies, say $S_1$ and $S_2$.  Consider, two optimization functions
\begin{enumerate}\itemsep 0mm
\item Minimize $(M_{T_1}(\vec{P}, \vec{S_1}, \vec{A}) \ast M_{E_1}(\vec{P}, \vec{S_1}, \vec{A}))$, subject to ${\cal F}_{T_1}(\vec{P}, \vec{S_1}, \vec{A}) \cap {\cal F}_{E_1}(\vec{P}, \vec{S_1}, \vec{A})$, and
\item Minimize $(M_{T_2}(\vec{P}, \vec{S_2}, \vec{A}) \ast M_{E_2}(\vec{P}, \vec{S_2}, \vec{A}))$, subject to ${\cal F}_{T_2}(\vec{P}, \vec{S_2}, \vec{A}) \cap {\cal F}_{E_2}(\vec{P}, \vec{S_2}, \vec{A})$
\end{enumerate}
As in \emph{(Auto) Super-Tuning}, we can formulate the problem of optimizing across strategies in two ways: (i) Take the minimum of the two optimizations, or (ii) solve separate optimization problems, depending on the intersections and differences of the feasible spaces of each one.  The methods can be extended to a set of strategies, ${\cal S} =\{ S_i\}$.

\end{enumerate}
Thus, our approach would allow us to deliver on the promise of \emph{automatic} and \emph{provably optimal} compilation, for any point in the program $\times$ transformations/strategies plane for a given performance metric.
\subparagraph{Polyhedral compilation revisited:}
As noted before, current polyhedral compilers target a fairly broad class of programs, and make choices like tiling hyperplanes and shapes, and (inter and intra) tile schedules.  They do this by this using classic scheduling algorithms~\cite{uday-pldi08, feautrier92a, feautrier92b} that use (integer) linear programming using surrogate objective functions.  Tile sizes are chosen subsequently via auto-tuning.

%On the other hand, we focus on a limited set of transformations/strategies, that can moreover be completely specified without such solvers.  Next, we formulate, for a given program and a given transformation strategy, an optimization problem that models execution time as a function of (i) the program parameters, including the dependence vectors, the iteration space bounds, and the number and size of data arrays, and (ii) the machine parameters.  We solve the problem to determine the optimal compilation choices (e.g., tile sizes).  Note that the solution to such an optimization problem may depend on program parameters, e.g., the parameters of the iteration space, potentially implying the need to solve separate optimization problem for each such size.  To overcome this, we would seek solutions that remain optimal for a range of problem sizes.

%We recently did this for the HHC strategy for GPUs~\cite{ppopp17} with excellent results.  Our models were within 10\% of the best performance on a set of ``validation'' data, and we were able to predict optimal tile sizes to further improved the performance.  We have developed similar analytic models for energy for GPUs~\cite{sanjay-energy-model-HPPAC2015}.  

\subsection{Codesign and its optimization subspaces}
\label{sec:codesignoptimization}

Codesign---the simultaneous design of hardware and software---has two common interpretations.  \emph{System codesign} is the problem of simultaneously designing hardware, runtime system, compilers, and programming environments of entire computing systems, typically in the context of large-scale, high-performance computing (HPC) systems and supercomputers. \emph{Application codesign}, also called \emph{hardware-software codesign}, is the problem of systematically and simultaneously designing a dedicated hardware platform and the software to execute a single application (program). The proposed approach is applicable in both contexts.

\subsubsection{Application Codesign and its optimization subspaces}
\begin{enumerate}
\item \textbf{Application Codesign:} 
%{Codesign: including architecture parameters in the optimization}
For \emph{Application Codesign}, we fix a program, a strategy as well as a micro-architecture (technology node).  In addition to the program parameters and features of the selected strategy, we now have $\vec{A}$, the architecture parameters, as unknowns.  The architecture parameters can be number of processors, number of memories, size of memories, etc.  Considering, execution time as the performance metric, the \emph{four-dimensional} optimization problem can be formulated as
\[ \mathrm{Minimize~~} M_{T}(\vec{P}, \vec{S}, \vec{A}), \mathrm{~~subject~to~~}{\cal F}_{T}(\vec{P}, \vec{S}, \vec{A}) \]
The feasible space ${\cal F}_{T}$ is a combination of (i) all possible program parameters, (ii) all possible features of the selected strategy, (iii) all possible architecture parameters, and (iv) time as the performance optimizing criteria.  Similar objective functions can be formulated for different performance metrics.

\item \textbf{Super Application Codesign:} 
Classic application codesign techniques consider optimization for one strategy.  Domain specificity allows us to formulate the mathematical optimization across a set of strategies, say ${\cal S} =\{ S_i\}$, as follows
\[ \mathrm{Minimize~~} M_{T}(\vec{P}, \vec{\cal S}, \vec{A}), \mathrm{~~subject~to~~}{\cal F}_{T}(\vec{P}, \vec{\cal S}, \vec{A}) \]
Note that the feasible space for each strategy will vary and not all architectures may be feasible for every strategy.  Such optimization allows for exploration of a larger feasible space.

\item \textbf{Multi-Metric Application Codesign:} 
The architecture parameters may include configuration options that trade off energy for performance (as in execution time).  One might, therefore, want to optimize across multiple performance metrics.  A multi-metric SW-HW Codesign optimization can be formulated as
\[ \mathrm{Minimize~~} (M_{T}(\vec{P}, \vec{S}, \vec{A}) \ast M_{E}(\vec{P}, \vec{S}, \vec{A})), \mathrm{~~subject~to~~}{\cal F}_{T}(\vec{P}, \vec{S}, \vec{A}) \cap {\cal F}_{E}(\vec{P}, \vec{S}, \vec{A}) \]
considering \emph{energy-delay product} as the performance metric.  Note that this \emph{four-dimensional} feasible space consists of program parameters, strategy features and architecture parameters as unknowns.

\item \textbf{Multi-Metric-Super Application Codesign:} 
Again, the multi-metric application codesign objective function can be extended to multiple strategies, say ${\cal S} =\{ S_i\}$.  Considering energy-delay product as the optimizing criteria, the objective function will be 
\[ \mathrm{Minimize~~} (M_{T}(\vec{P}, \vec{\cal S}, \vec{A}) \ast M_{E}(\vec{P}, \vec{\cal S}, \vec{A})), \mathrm{~~subject~to~~}{\cal F}_{T}(\vec{P}, \vec{\cal S}, \vec{A}) \cap {\cal F}_{E}(\vec{P}, \vec{\cal S}, \vec{A})\]
Note that the feasible space here consists of multiple areas in the design space based on the strategy and performance metric.
\end{enumerate}

\subsubsection{System Codesign and its optimization subspaces}

\begin{enumerate}
\item \textbf{System Codesign:} 
Let us now consider a set of program instances ${\cal P} = \{ P_i \}$, and recall that we have \emph{common} objective functions for all of them.  The optimization problem
\[ \mathrm{Minimize~~} {\cal M}({\vec{\cal P}}, \vec{S}, \vec{A}) \mathrm{~~subject~to~~} {\cal F}(\vec{{\cal P}}, \vec{S}, \vec{A}) \]
seeks to optimize a common performance metric (we drop the subscript for the common metric).  The parameters to this function are a set of program instances, features of the strategy and the architecture features.

\item \textbf{Super System Codesign:} 
Let us now consider the following optimization problem, which we call the \textbf{\emph{generalized optimization problem}}.
\[ \mathrm{Minimize~~} {\cal M}({\vec{\cal P}}, \vec{{\cal S}}, \vec{A}) \mathrm{~~subject~to~~} {\cal F}(\vec{{\cal P}}, \vec{\cal S}, \vec{A}) \]
This problem seeks to optimize the common performance metric for the set of programs on the given architecture.  We treat $\vec{A}$, not as parameters, but rather as unknowns, in the generalized optimization problem, $\displaystyle \mathrm{argmin}_{\vec{A}} {\cal M}(\vec{\cal P}, \vec{\cal S}, \vec{A})$ gives us the optimal architecture for the set of program instances.  Thus we simultaneously solve for architecture and compilation, thereby resolving the \emph{\textbf{codesign problem}}.  %An early vision of this was articulated by Djidjev~\cite{hristo-TR} and we have been collaborating with him on this for the past few years.

\item \textbf{Multi-Metric System Codesign:} 
The next step would be to extend our \textbf{\emph{generalized optimization problem}} to a multi-metric optimization. 
\[ \mathrm{Minimize~~} (M_{T}({\vec{\cal P}}, \vec{S}, \vec{A}) \ast M_{E}({\vec{\cal P}}, \vec{S}, \vec{A})), \mathrm{~~subject~to~~} {\cal F}_{T}(\vec{{\cal P}}, \vec{S}, \vec{A}) \cap {\cal F}_{E}(\vec{{\cal P}}, \vec{S}, \vec{A}) \]
This is particularly useful in large system designs, where the transformation strategy is fixed and more than one performance metric is critical for system design.  Note, that we show multi-metric optimization for two cost metrics which can be extended to more than two cost metrics as needed.

\item \textbf{Multi-Metric-Super System Codesign:} 
The above multi-metric system codesign can be further extended to consider multiple strategies as shown below
\[ \mathrm{Minimize~~} (M_{T}({\vec{\cal P}}, \vec{\cal S}, \vec{A}) \ast M_{E}({\vec{\cal P}}, \vec{\cal S}, \vec{A})), \mathrm{~~subject~to~~} {\cal F}_{T}(\vec{{\cal P}}, \vec{\cal S}, \vec{A}) \cap {\cal F}_{E}(\vec{{\cal P}}, \vec{\cal S}, \vec{A}) \]
This would be the \emph{\textbf{ultimate goal for system codesign}} where we can optimize across all the possible program, transformation and architecture planes for multiple performance metrics.

\end{enumerate}
\subsection{Bottleneckology}
\label{sec:approachbottleneckology}
Our approach to system design seems deceptively simple, however, it is a very hard problem.  Exploiting the resources to their full capacity is one of the objectives while optimizing for performance.  The bottleneck analysis becomes helpful in studying the performance sinks and design flaws.  There are many ways of utilizing the cost models to perform bottleneck analysis.  The cost models can be used to identify the resources that have been saturated and the ones that have slack.  We refer to this slack and saturation of the resources as \emph{\textbf{Bottleneckology}}.  We study this in three ways: (i) investigate codesign-tradeoffs, (ii) perform overhead analysis, and (iii) explore the effect of hyperthreading.  More details are provided in Chapter~\ref{chap:bottleneckology}.

In the next chapters (\ref{chap:costmodels},~\ref{chap:autotuning},~\ref{chap:codesign}, and~\ref{chap:bottleneckology}), we discussed our work in more details.
%
%
% Local Variables: ***
% TeX-master: "CHACK.tex" ***
% fill-column: 78 ***
% End: ***

% CHACK - design space + energy
% Lakshmi's thesis
% Research Objectives - THESIS STATEMENT

\chapter{Models and Validation}
\label{chap:costmodels}
%%%%%%%%%%%%%%%%%%%%%%%%%%%%%%%%%%%%%%%%%%%%%%%%%%%%%%%%%%%%%%%%
%
Model predictions are used for estimating execution time, energy consumption, power consumption, etc. of a program.  Cost metrics either appear in the objective function as a factor to be optimized or in the constraints.  We will illustrate the use of a few of the many metrics - (i) execution time models, and (ii) energy models as cost in the optimizing functions; and (iii) memory access models, and (iv) silicon area models as the constraints to the objective function.
%
%
%%%%%%%%%%%
%%%   Time   %%%
%%%%%%%%%%%
\section{Execution Time Model for GPGPU Stencils}
\label{sec:timemodel}
%%%%%%%%%%%
%
%Stencil computations are an important class of compute and data intensive programs that occur widely in scientific and engineering applications.  A number of tools use sophisticated tiling, parallelization, and memory mapping strategies, and generate code that relies on vendor-supplied compilers.  This code has a number of parameters, such as tile sizes, that  are then tuned via empirical exploration. 
We develop an execution time models for GPGPU stencils that guides the optimal choice of compiler parameters(tile sizes).  Our model is a simple set of analytical functions that predict the execution time of the generated code. It is deliberately optimistic, since we are targeting modeling and parameter selections yielding highly optimized codes. We experimentally validate the model on a number of 2D and 3D stencil codes, and show that the root mean square error in the execution time is less than 10\% for the subset of the codes that achieve performance within 20\% of the best.  We show the following.
\begin{itemize}\itemsep 0mm
\item We develop a simple analytical model to predict the execution time of a tiled stencil program and apply it to codes generated by the HHC compiler. The model is an analytic function of
  \begin{itemize}\itemsep 0mm
  \item program, machine, and compiler parameters that are easily available statically, and
  \item one stencil-specific parameter that is obtained by running a handful of \emph{micro-benchmarks}.
  \end{itemize}
  It is deliberately \emph{optimistic} and also ignores the effect of some parameters.
\item Although our model may not accurately predict the performance for all tile size combinations, it is very accurate for the ones that matter, i.e., those that give top performance.  To show this, we generated more than than 60,000 programs for
   \begin{itemize}
   \item two modern \textbf{target platforms} (NVIDIA GTX~980, and Titan~X), and
   \item four 2D \textbf{stencil codes} (Jacobi2D, Heat2D, Laplacian2D, and Gradient2D) and two 3D stencils (Heat3D and Laplacian3D)
   \item over a range of ten input/problem \textbf{sizes}, and
   \item a wide range of tile sizes and thread counts (the HHC compiler inputs) for each \emph{platform-stencil-size} combination. 
   \end{itemize}
   As we expected, the root-mean-square error (RMSE) over the \textbf{entire} data set was ``disappointingly'' over 100\%.  However, when we restricted ourselves to the data points that have an execution time within 20\% of the best value for that particular platform-stencil-size combination, the RMSE dropped to less than 10\%\footnote{The restriction to the better performing subset was exactly our motivation.  We designed the model to help predict/explore data points that would give \emph{good} performance.  It is also why we made optimistic assumptions in developing the model.}, which we consider very good.
\end{itemize}

Our overall methodology is applicable, with simple extensions, to more general programs, e.g., those that fit the polyhedral model.  But for achieving high GPU utilization, we need efficient GPU codes to start with, which are very hard and time consuming to produce manually, especially in higher dimensions. The highly optimized HHC-generated codes we are using for testing and validation have a few thousand lines of CUDA code each and we generated tens of thousands such codes in our experimental analysis.  So our methodology is not limited to the HHC compiler (in fact we have applied it successfully to manually generated 1D stencil codes), but the use of HHC (or similar compiler) was necessary to produce for our experiments a high number of GPU codes that are also very efficient.
%
%
%%%%%%%%%%%
%%%  Energy  %%%
%%%%%%%%%%%
\section{Energy Model for Tiled Nested-Loop Codes}
\label{sec:Energy}
%%%%%%%%%%%%%%%%%%%%%%%%%%%%%%%%%%%%%%%%%%%%%%%%%%%%%%%%%%%%%%%%
%
Energy efficiency has been recognized as one of the biggest challenges in the roadmap to higher performance (exascale) systems for a number of reasons including cost, reliability, energy conservation, and environmental impact.  The most powerful computers today consume megawatts of power, enough to power small towns, and at cost of millions per year. And those estimations do not include the cost of cooling, which might be almost as high as the cost of computing itself \cite{Rajamani2003}.  In addition, the cost of building a power provisioning facility ranges at \$10-22 per deployed IT watt \cite{Barroso2009} and every $10\,^{\circ}\mathrm{C}$ temperature-increase results in a doubling of the system failure rate, thereby reducing the reliability of HPC systems \cite{Rodero2012}. Designing accurate models for energy efficiency can help better predict the power and energy requirements of an application and aid developers optimize the parameters of their codes for better energy efficiency on HPC systems.

The goal of our work is to introduce a new approach for modeling the energy cost as an analytical function of tunable software parameters in a way that is both simple and accurate. Having such a model will allow the energy efficiency to be optimized with respect to (a subset of) the tunable parameters by solving the corresponding analytical optimization problem. 

We target with our modeling approach tiled nested-loop code segments, which are the most compute-intensive portions of many application codes and which also allow a high degree of parallelism. In order to be more specific, we focus in our analysis on a subclass of the tiled nested-loop codes called \emph{dense stencils}, which occur frequently in the numerical solution of PDEs and in many other contexts such as high-end graphics, signal and image processing, numerical simulation, scientific computing, and bioinformatics. We chose stencils  for our case studies since that would allow us to model the entire class in a hierarchical way with a single generic model representing the whole class, while model parameters that are stencil-dependent have to be separately specified for each stencil of interest to complete its model.  (However, the approach is applicable to any other class of nested-loop codes that allows tiling.) We completely develop and validate the detailed models (including the stencil-dependent parameters) of three specific stencils. Models for other stencils can be developed in a similar way with relatively small amount of extra work.  

In order to efficiently optimize stencils on accelerators, we aim to represent the amount of energy consumed as an analytic function of the software parameters.  We assume that the input codes have been analyzed and optimized with respect to parallelism and data-access efficiency by appropriate skewing and tiling transformations, say by a polyhedral code generator.

Our specific contributions are as follows.
\begin{itemize}
	\item Our energy model predicts energy efficiency by analyzing source code only, unlike other approaches \cite{Nagasaka10, Leng13:GPUWattch} that rely on parameters computed by running benchmarks for each individual code.  We do use micro-benchmarks, but they are used to characterize hardware, rather than codes.
	\item We are not aware of any previous work combining the polyhedral method with energy modeling.  Our approach allows optimization of codes that are already very efficient having been significantly improved by applying the polyhedral method and by using advanced tiling strategies such as hexagonal and hybrid tilings \cite{grosser-etal-GPUhextile-CGO2014}. 
	\item Our model is very accurate (one version with $\mathrm{RMS~error} \leq 17.14\%$  and another with $\mathrm{RMS~error} \leq 4\%$), with similar or higher precision than alternative existing models, e.g., GPUSimPow~\cite{gpusimpow}, which are simulation based. %, i.e., require (at least partial) execution.
\end{itemize}
%
%
%%%%%%%%%%%
%%    Memory   %%
%%%%%%%%%%%
\section{Memory Access Model for GPGPU Stencils}
\label{sec:memorymodel}
%%%%%%%%%%%
We develop Memory Access models~\cite{prajapati2017simple, prajapati-ipdpsw-energy} for GPGPU stencils and use them for execution time models and energy models.  The memory models appear in two specific contexts.  Firstly,  the total number of memory accesses made by a tile is used to model the data transfer time taken by a tile, similarly, the data movement requirement of a wavefront is modeled using the equations to calculate the data transfer time for wavefronts.  Similarly, for the energy models we use memory access equations to determine the amount of transfer required and combine it with the energy consumption per data transfer to calculate the total energy consumption of a tile.

Second, the memory footprint of a tile appear as constraints to the formulated objective function for optimization.  The memory requirements of a tile are not to exceed the shared memory capacity of a GPU.  This in turn constrains the tile sizes and the feasible space.

These memory models are then used for codesign optimization~\cite{prajapati2017techreport}.  Again for software-hardware codesign, the memory models appear both in the objective function as a part of time model equations (i.e. to calculate the data transfer time) as well as the constraints where memory capacity defines the feasible space.

\section{Silicon Area Model for GPUs}
\label{sec:areamodel}
%%%%%%%%%%%%%%%%%%%%%%%%%%%%%%%%%%%%%%%%%%%%%%%%%%%%%%%%%%%%%%%%
%
We develop an analytic model for the total silicon area of a GPU accelerator.  We faced some difficulties in deriving an acceptable analytical model, as silicon data had to be reverse engineered from extremely limited public domain resources.  As a general observation, within each GPU family, there is little diversity in the parameter configurations.  For the Maxwell family of GPUs, the GTX980 and Titan~X chips were chosen as two sufficiently distinct points to calibrate our analytical models.  The calibration itself was performed by evaluating die photomicrographs, publicly available information about the nVidia GTX-980 (Maxwell series) GPU, and other generally accepted memory architecture models.  The model validation was done by comparing the predictions with known data on the Maxwell series Titan X GPU. We found the model prediction to be accurate to within, 2\%, though this number is not significant.\footnote{Although a many configurations of any family of GPUs are spaced out, they come from binning only a small number of distinct dies.  We ended up calibrating our model on one die and validating it on only another one.}

In the next chapter, we will show the use of these cost models for tile size selection.
% Time
% Energy
% Validation
%
% Show Problem Formulation and Design Space
% Questions
%    i. What happens when input parameters change?
%    ii. When different number of processors is used?
%    iii. Largest possible problem size
%    iv. When does efficiency drop? 

%\include{chap5-AreaModel}
% Area
% Validation
%
% Show Problem Formulation and Design Space
% Talk about how it can be used for system design
% System Tuning: tune nvcc parameters, cuda installation, runtime

\chapter{Tuning}
\label{chap:autotuning}
%%%%%%%%%%%%%%%%%%%%%%%%%%%%%%%%%%%%%%%%%%%%%%%%%%%%%%%%%%%%%%%%

An important element of compilation tools is a step called \emph{(auto) tuning}: empirical evaluation of the actual performance of a, hopefully small, set of code instances for a range of mapping parameters.  This enables the compilation system to choose these parameters optimally for actual ``production runs'' on real data/inputs.  Modern architectures are extremely complicated, with sophisticated hardware features that interact in unpredictable manners, especially since the latency of operations is unpredictable because of the deep memory hierarchy.  It is widely believed that because of this, autotuning is unavoidable in order to obtain good performance.

Our work challenges this.  In particular, we make the case that domain specificity can have a third important benefit: it enables us to develop a good analytical model to predict the performance of specific types of codes on specific types of target architectures.  We can then use the model to optimally choose the mapping parameters (notably tile sizes).  

In order to address the challenges of exascale computing, many experts believe that a software-hardware co-design approach---where the software and the corresponding hardware are jointly co-developed and co-optimized---will be a ``critical necessity'' \cite{Sarkar09}.  Since the architectures of exascale systems are in the flux, it is important to develop rigorous methods to map high level specifications of computations to diverse target architectures, ranging from multi-core CPUs, many-core GPUs, and accelerators over heterogeneous nodes of such CPU-GPU combinations to large distributed systems of many such nodes.  In the overwhelming majority of cases, the mismatch between data communication patterns and hardware architecture prevents the efficient exploitation of all available computing resources and peak performance is almost impossible to achieve.  Worse still, it is often not clear to the user when the point of diminishing returns is reached.

We address a key step of the optimization, namely mapping the software representation onto the hardware, and choosing the mapping parameters to optimize an objective function representing the performance,  i.e., the execution time.  In its full generality, the optimal mapping problem is a discrete non-linear optimization problem, known to be NP-hard~\cite{Ullman:1975} and hence very difficult to solve efficiently. We therefore use a number of simplifying assumptions, as is common in the literature. A number of parameters can be specified as inputs to a compiler, e.g., the tile sizes. These parameters have a tremendous influence on the performance of the code. The problem we tackle here is how to select these parameters optimally.  

%Time
\section{Tune for Speed}

To test the predictive abilities of our execution time models, we evaluated the model over the entire feasible space (for each platform-stencil-size combination) and obtained the tile sizes that were within 10\% of the best predicted execution time.  There were less than 200 such points.  We called the HHC compiler with these tile sizes and were able to observe among this set a performance improvement of 9\% on average with maximum of 17\%.  Prajapati et al.~\cite{prajapati2017simple} illustrates the predictive power of our execution time models in detail.

We have two messages.  The main one is that, contrary to widespread belief, it is possible to construct good analytical cost functions to drive performance tuning for GPGPUs.  This can significantly reduce the space that autotuners need to explore.  The second message, is that it may be necessary to revisit some of the ``conventional wisdom,'' when choosing tile size parameters.  Our model is very accurate for predicting the times of problem instances whose performance is within 20\% of the optimal and, hence, it can be used to find values for tunable parameters that will give near optimal performance. 

We would like to note that our techniques can be easily extended to other type of stencils.

%We applied our model for optimizing the tile sizes and experimentally observed a noticeable improvement in performance when compared with manually determined best tile sizes found after significant numbers of experiments.

%While exploring the predictive capability of the model, we explored all points with predicted performance within 10\% of $T_\mathrm{Alg\_min}$.  This is because our model does not account for many architectural and code features: thread divergence, imbalance among threads in the same warp, branch divergence, memory bank conflicts, etc.  We also do not model the effect of the number of registers
%% Moreover, the number of thread blocks that can simultaneously execute on an
%% $SM$, $K$, is dependent on two factors $K_\mathrm{smem}$ and
%% $K_\mathrm{reg}$.
%per thread block, a factor that can only be obtained ``post mortem'' after the \texttt{nvcc} compiler.  This is why it is still necessary to have an empirical tuning phase, but we have shown that the number of points that need to be explored in this manner is relatively small.

% Energy
\section{Optimize for Energy}
%We also show that the models can be used for optimal tile-size selection for energy efficiency.  
% Our proposed optimization method will then additionally optimize the codes
% with respect to the tile sizes.
Our proposed optimization methods can also be applied to optimize for energy. Our energy models represents the energy consumption as an explicit \textit{analytic} function of a set of software and hardware parameters describing the specifics of the implementation and the hardware.  %We are not aware of any similar result. Moreover, while we use our model here only for tile size selection, it may be used in other types of scenarios, such as codesign (simultaneously optimizing the hardware and software parameters). % We are planning to look for such applications in future works.

In our experiments, we observe that energy optimization has almost always had no loss of speed, as expected in the folklore.  Thus a user could use both optimizations rather than having to choose one or the other.  Finally, we use our energy model to select the optimal tile size for energy efficiency and report the number of non-optimal tile size selections and hence the error in energy due to the selection of non-optimal tile sizes.  Prajapati et al.~\cite{prajapati-ipdpsw-energy} describes our energy models, the optimization methods, the results and experimental validation in details. 
%\input{HPPAC/sections/optimizing-energy}

%We believe that the approach we developed can be adapted and applied to other related problems. One extension is to combine the energy model with an execution-time models. We can also apply the approach to a wider class of nested-loop codes that includes but is not restricted to stencil codes and to platforms other than GPU. We can also model and optimize with respect to the maximum power instead of energy, or extend the model to allow dynamic voltage and frequency scaling optimization. 

% Performance Improvement
% i. tile size optimization, and ii. energy optimization
% Application Design---??
%
% AutoSuperTuning :) --- hardware parameters are fixed, how to do tune for compiler parameters. For eg. loop skewing, loop fusion, tiling, etc. which is the best transformation? 
% Formulate the problem and show design space. GOOD IDEA
%
% Time+Energy co-optimization
% Formulate the problem and show design space. 

\chapter{Accelerator Codesign}
\label{chap:codesign}
%%%%%%%%%%%%%%%%%%%%%%%%%%%%%%%%%%%%%%%%%%%%%%%%%%%%%%%%%%%%%%%%

% ICS paper - our codesign

\emph{``Design is not just what it looks like and feels like. Design is how it works.''}  -- Steve Jobs

\mbox{}
%We propose an optimization approach for determining both hardware and software parameters for the efficient implementation of a (family of) applications called \emph{dense stencil computations} on programmable GPGPUs.  We use our analytical model for the silicon area usage of accelerator architectures and a workload characterization of stencil computations.  We combine this characterization with our parametric execution time model and formulate a mathematical optimization problem.  That problem seeks to maximize a common objective function of \emph{all the hardware and software parameters}.  The solution to this problem therefore ``solves'' the codesign problem: simultaneously choosing software-hardware parameters to optimize total performance.
%
%We validate this approach by proposing architectural variants of the NVIDIA Maxwell GTX-980 (respectively, Titan~X) specifically tuned to a predetermined workload of four common 2D stencils (Heat, Jacobi, Laplacian, and Gradient) and two 3D ones (Heat and Laplacian).  Our model predicts that performance would potentially improve by 28\% (respectively, 33\%) with simple tweaks to the hardware parameters such as adapting coarse and fine-grained parallelism by changing the number of streaming multiprocessors and the number of compute cores each contains.  We propose a set of Pareto-optimal design points to exploit the trade-off between performance and silicon area and show that by additionally eliminating GPU caches, we can get a further 2-fold improvement.

Software-hardware codesign is one of the proposed enabling technologies for exascale computing and beyond~\cite{b1282de6ab8a4f1a8a962cb40310bc7f}. Currently, hardware and software design are done largely separately.  Hardware manufacturers design and produce a high-performance computing (HPC) system with great computing potential and deliver it to customers, who then try to adapt their application codes to run on the new system.  But because of a typically occurring mismatch between hardware and software structure and parameters, such codes are often only able to run at a small fraction of the total performance the new hardware can reach.  Hence, optimizing both the hardware and software parameters \textit{simultaneously} during hardware design is considered as a promising way to achieve better hardware usage efficiency and thereby enabling leadership-class HPC availability at a more manageable cost and energy efficiency.

The design of HPC systems and supercomputers is by no means the only scenario where such optimization problems occur.  The execution platforms of typical consumer devices like smart phones and tablets consist of very heterogeneous Multi-Processor Systems-on-Chip (MPSoCs) and the design challenges for them are similar.

Despite the appeal of an approach to \emph{simultaneously} optimize for software and hardware, its implementation represents a formidable challenge because of the huge search space.  Previous approaches
\cite{codesign-white-paper, opt_principles, Mniszewski:2015:TDE:2764453.2699715}, pick a hardware model ${\cal H}$ from the hardware design space, a software model ${\cal S}$ from the software design space, map ${\cal S}$ onto ${\cal H}$, estimate the performance of the mapping, and iterate until a desirable quality is achieved. But not only each of the software and hardware design spaces can be huge, each iteration takes a long time since finding a good mapping of ${\cal S}$ onto ${\cal H}$ and estimating the performance of the resulting implementation are themselves challenging computational problems.

We propose a new approach for the software-hardware codesign problem that avoids these pitfalls by considerably shrinking the design space and making its exploration possible by formulating the optimization problem in a way that allows the use of existing powerful optimization solvers.  We apply the methodology to programmable accelerators: Graphics Processing Units (GPUs), and for stencil codes.  The key elements of our approach are to exploit multiple forms of \emph{domain-specificity}.
Our main contributions are:
\begin{itemize}
\item We propose a new approach to software-hardware codesign that it is computationally feasible and provides interesting insights.
  % allows finding designs nearly optimal with respect to both software and
  % hardware parameters
\item We combine our area model with a workload characterization of stencil codes, and our previously proposed execution time model~\cite{prajapati2017simple} to formulate a mathematical optimization problem that maximizes a common objective function of the hardware and software parameters.
\item Our analysis provides interesting insights.  We produce a set of Pareto optimal designs that represent the optimal combination of hardware and compiler parameters.  They allow for up to $33\%$ improvement in performance as measured in GFLOPs/sec.
\end{itemize}

We develop a framework for software-hardware codesign that allows the simultaneous optimization of software and hardware parameters. It assumes having analytical models for performance, for which we use execution time, and cost, for which we choose the chip area. We make use of the execution time model from Prajapati et al~\cite{prajapati2017simple} that predicts the execution times of a set of stencil programs. For the chip area, we develop an analytical model that estimates the chip area of parameterized designs from the Maxwell GPU architecture. Our model is reasonably accurate for estimating the total die area based on individual components such as the number of SMs, the number of vector units, the size of memories, etc.

We formulate a codesign optimization problem using the time model and our area model for optimizing the compiler and architecture parameters simultaneously.  We predict an improvement in the performance of 2D stencils by (104\% and 69\%) and 3D stencils by (123\% and 126\%) over existing Maxwell (GTX980 and Titan~X) architectures. 

The main focus is on the methodology; specifically, to develop a software-hardware codesign framework and to illustrate how models built using it can be used for efficient exploration of the design space for identifying Pareto-optimal configurations and analyzing for design tradeoffs. The same framework, possibly with some modifications, could be used for codesign on other type of hardware platforms (instead of GPU), other type of software kernels (instead of the set of stencils we chose, or even non-stencil kernels), and other kind of performance and cost criteria (e.g., energy as cost). Also, with work focused on the individual elements of the framework, the execution time and the chip area models we used could possibly be replaced by ones with better features in certain aspects or scenarios. 

The analyses from our work indicate the following accelerator design recommendations, for the chosen performance, cost criteria, and  application profile: 
\begin{itemize}
\item Remove caches completely and 
\item Use the area (previously devoted to caches) to add more cores on the chip.
\item The more precise the workload characterization and the specific area model parameters, the more useful the conclusions drawn from the study.
\end{itemize}

Hardware resources such as memories are often expensive and must be utilized wisely.  In the next chapter, we discuss how to use cost models to identify the resources requirements for optimal performance via bottleneck analysis.

% SPU - how to design new architectures using simulator
% FPGAs - max+ design

% ICS paper - our codesign
% SPU - how to design new architectures using simulator
% FPGAs - max+ design
%
% include Energy - Show Problem Formulation and Design Space

\chapter{Bottleneckology}
\label{chap:bottleneckology}
%%%%%%%%%%%%%%%%%%%%%%%%%%%%%%%%%%%%%%%%%%%%%%%%%%%%%%%%%%%%%%%%

% how to show bottleneck analysis using time/energy/area/codesign
% Slack and Saturation of resources

We explore \emph{\textbf{bottleneckology}} in three ways: \emph{\textbf{study codesign trade-offs, perform overhead analysis, and investigate the effect of hyperthreading}}.  Next three sections discuss these in more details.
\section{Codesign Trade-offs}
\begin{table}
  \centering
  \begin{tabular}{||l||r|r|r|r|r||}
  \hline
    \multicolumn{1}{||c||}{Code} &
    \multicolumn{1}{|c}{$n_\mathrm{SM}$} &
    \multicolumn{1}{|c}{$n_\mathrm{V}$} &
    \multicolumn{1}{|c}{$M_\mathrm{SM}$} &
    % \multicolumn{1}{|c}{$\mathrm{NV*NSM}$} &
    % \multicolumn{1}{|c}{$\mathrm{MSM*NSM}$} &
    \multicolumn{1}{|c|}{Area} &
    \multicolumn{1}{|c||}{GFLOPs/S} \\ \hline\hline
    Jacobi 2D & 32 & 128 & 24 & 438 & 2059 \\
    Heat 2D & 22 & 256 & 12 & 447 & 3017 \\
    Gradient 2D & 28 & 160 & 24 & 431 & 4963 \\
    Laplacian 2D & 28 & 160 & 12 & 426 & 2549 \\
    Heat 3D & 18 & 288 & 192 & 447 & 3600 \\
    Laplacian 3D & 8 & 896 & 96 & 446 & 1427 \\
    \hline
  \end{tabular}
  \vspace*{2mm}
  \caption{Workload sensitivity. The optimal architecture configuration for a single benchmark varies significantly.}
  \label{tab:configurations}
\end{table}
\paragraph{Workload Sensitivity}
%
%We hypothetically set the frequency for one of the benchmarks as one (and zero for the others) thereby allowing us to explore designs optimized for a single kernel.  It also helps determine whether the chosen suite is representative of the mix that occurs in practice. 
Table~\ref{tab:configurations} illustrates the architectural parameters for the best performing designs for each of the six benchmarks(2D and 3D stencils), for an area budget between 425--450 $\mathrm{mm}^2$.  Observe how the parameters of the best architecture are significantly different.  There are also differences in the achieved performance for each benchmark, but that is to be expected since the main computation in the stencil loop body has different number of operations across the benchmarks. 
\paragraph{Shared Memory Requirements}
We can also observe that there are marked differences between the optimal architecture configurations for 2D and 3D stencils in Table~\ref{tab:configurations}.  3D stencils seem to require larger shared memory ($\geq 96$ kB / SM) compared to 2D stencils ($\leq 24$ kB / SM).  Indeed, for designs with lower than 48kB, the performance was nowhere near the optimal for 3D stencil programs.  Comparing the optimal configurations for $Heat~2D$ stencil with that of $Heat~3D$ stencil (both have equal total die area of $447$ $mm^{2}$), we observe that the amount of shared memory required for $Heat~3D$ stencil is $16$ times more than that for $Heat~2D$ stencil.  Also note, 3D stencils require higher number of vector units per SM for optimal performance. 
\begin{figure*}[tb]
\centering
\includegraphics[width=3in]{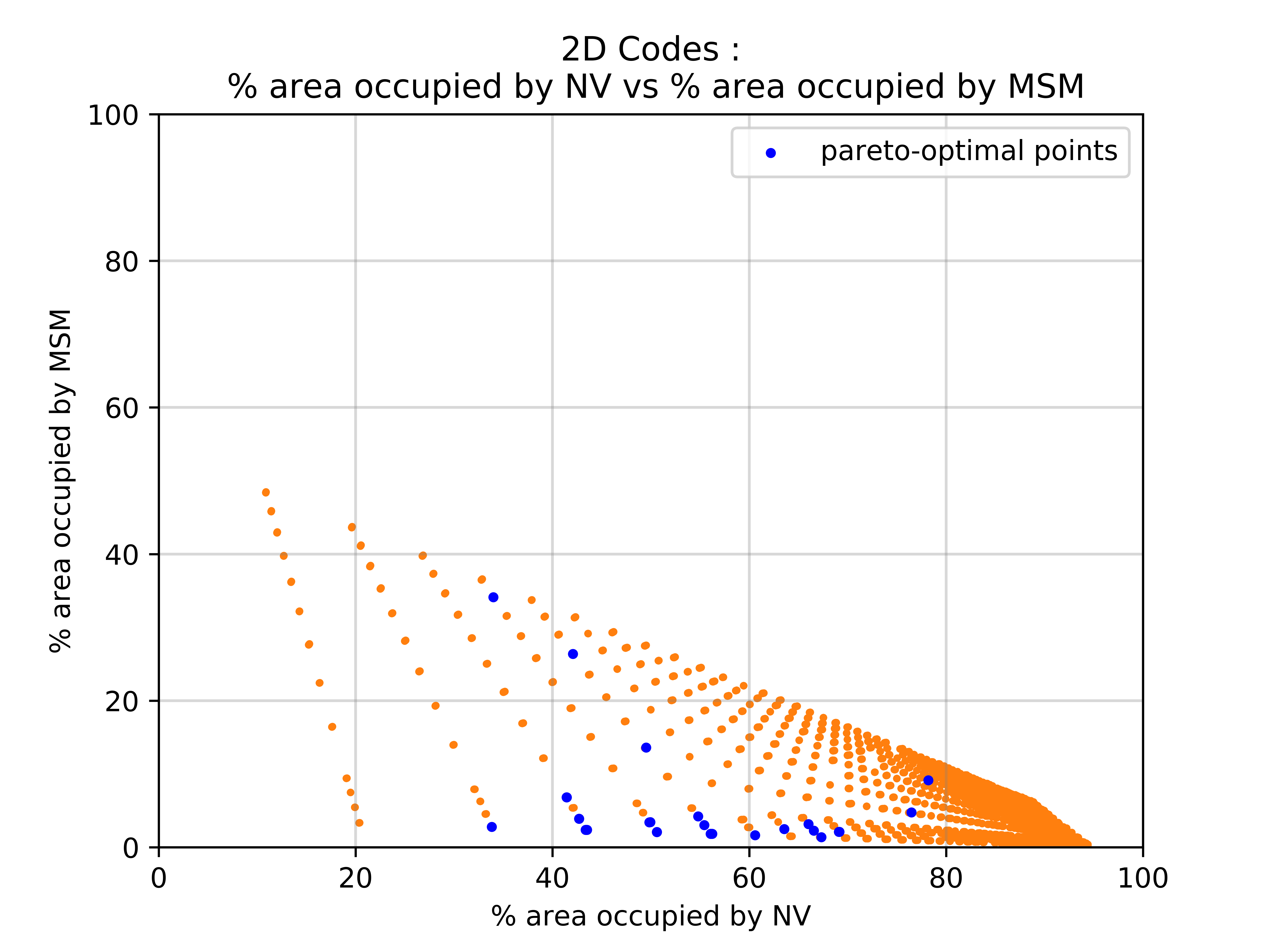}
\includegraphics[width=3in]{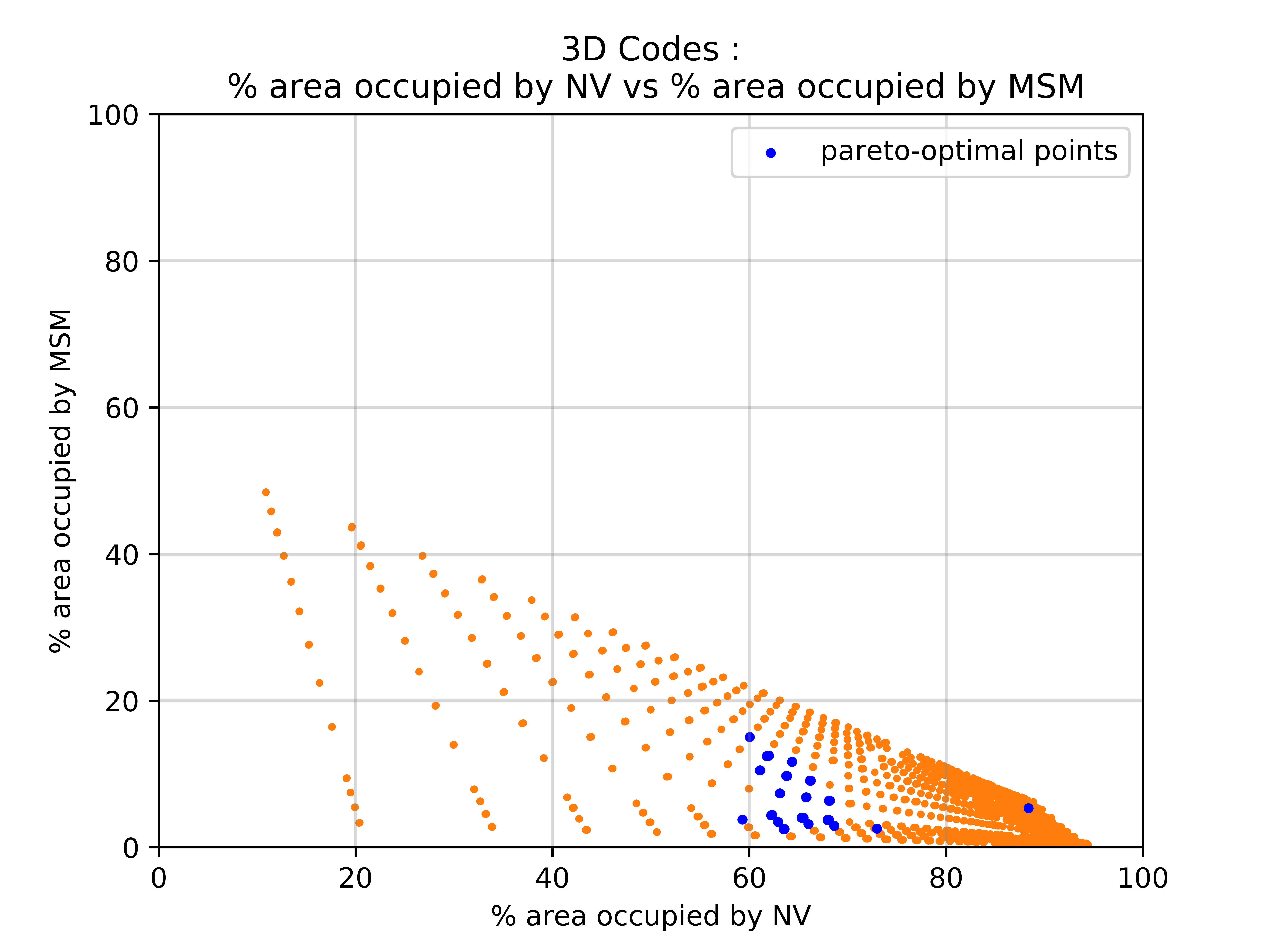}
\protect\caption{Resource Allocation.}
\label{fig:resource}
\end{figure*}
\paragraph{Resource Allocation}
Another interesting perspective is seen in Figure~\ref{fig:resource} which plots the Pareto-optimal design points in blue and all other non-Pareto configurations in orange.  The axes show the relative percentages of the chip area devoted to memory, and to vector units.  We notice that the optimal designs (blue points) lie in a relative cluster.  This phenomenon is even more marked for 3D stencils. At present, we do not have a clear explanation for why the points are clustered in this manner, and we plan to mine this data to determine patterns, if any.

\section{Overhead Analysis}
%\section{Overhead Analysis and Pruning the Tile Size Design Space}

%Stencil computations are an important class of compute and data intensive programs that occur widely in scientific and engineering applications. 
%A number of tools use sophisticated tiling, parallelization, and memory mapping strategies, and generate code that relies on vendor-supplied compilers. 
As discussed before, a compiler generated code has a number of parameters, such as tile sizes, that are then tuned via empirical exploration. Our execution time model~\cite{prajapati2017simple} guides such a choice. This execution time model is a simple set of analytical functions that predicts the execution time of the HHC~\cite{grosser-etal-GPUhextile-CGO2014} generated code. 

The execution time model uses a machine dependent parameter, called $C_\mathrm{iter}$, which is the execution time of one iteration of the loop body per vector unit provided that all the necessary data is available in shared memory.  To measure $C_\mathrm{iter}$, one needs to generate a random set of codes for a given stencil with different tile sizes, modify these codes to remove global memory accesses and then take the average of empirically measured execution times to obtain the value of $C_\mathrm{iter}$.  The process is very time consuming and requires expertise in Hybrid Hexagonal Tiled code generator~\cite{grosser-etal-GPUhextile-CGO2014}.  Also, $C_\mathrm{iter}$ is dependent on both, the machine as well as the program.  Therefore, its value changes as machine and program parameters vary.  It is, therefore, very difficult to model/measure is $C_\mathrm{iter}$.   Also note that $C_\mathrm{iter}$ value is used to evaluate the objective function to find the optimal tile size.  This imposes limitations on the use of execution time model for optimal tile size selection because the crucial model parameter $C_\mathrm{iter}$, is to be empirically measured.

To address this problem, we propose a closed form solution that is completely independent of the machine and the program parameters.  We are able to analytically predict the optimal tile sizes which are portable across platforms and is valid for all Jacobi-like stencils.  We modify the objective function from Prajapati et al.~\cite{prajapati2017simple} and develop a cost function which is independent of $C_\mathrm{iter}$.  For a 2D stencil, our closed form solution suggests that we maximize the size of the hexagonal face of a tile subject to some constraints. This allows us to significantly narrow down the tile size design space.

The mathematical optimization problem in Prajapati et al.~\cite{prajapati2017simple} for a given 2D stencil is formulated as follows:

\begin{equation}\label{optimize}
minimize  \;\;\; T_{alg}(t_{S_{1}}, t_{S_{2}}, t_{T})
\end{equation}

where $t_{S_{1}}$, $t_{S_{2}}$, and $t_{T}$ are tile sizes and $T_{alg}$ is the model predicted execution time of the code given by the following formula:

\begin{equation}\label{t_alg_2D}
  T_\mathrm{alg} = N_w T_\mathrm{sync} + N_w T_\mathrm{prism} \left\lceil
    \frac{1}{n_{SM}} \left\lceil \frac{w}{k} \right\rceil\right\rceil.
\end{equation}

where $N_w$ is the number of wavefronts, $T_\mathrm{sync}$ is the time for synchronization for a wavefront, $T_\mathrm{prism}$ is the time to execute a tile, $n_{SM}$ is the number of processors, $w$ is the size of a wavefront and $k$ is the number of tiles that execute simultaneously. For more details, please refer Prajapati et al.~\cite{prajapati2017simple}.

In addition to the time for computation, $T_{alg}$ includes time taken by data transfers and the time for inter-tile and intra-tile synchronizations.  We are interested in only those tile sizes that give optimal performance.  Therefore, our tiles will be compute bound.  Let us consider an ideal machine where the performance is given by the following equation:

\begin{equation}\label{t_ideal}
  T_\mathrm{ideal} = \frac{S_1 S_2 T}
    {n_{SM} n_{V}} C_{iter}
\end{equation}

where $S_1$, and $S_2$ are problem sizes in space dimensions, $T$ is the problem size in time dimension, and $n_V$ is the number of vector units.  Such an ideal machine is free of all synchronization delays and takes no additional time to do data transfers. On a real machine, we would like to obtain the performance that is close to $T_{ideal}$.  However, there is always an overhead price such that 

\begin{equation}\label{t_ov}
  T_\mathrm{overhead} = T_\mathrm{alg} - T_\mathrm{ideal}
\end{equation}

Substituting $T_\mathrm{alg}$ and $T_\mathrm{ideal}$ with their respective equations and solving gives us

\begin{equation}\label{t_ov_2}
  T_\mathrm{overhead} = C_{iter} S_1 T \frac{1} {t_{S_1} + \frac{t_{T}}{2}} 
\end{equation}

Instead of minimizing the execution time(as in the equation~\ref{optimize}), we can now minimize the overhead.  To minimize equation~\ref{t_ov_2}, we need to maximize ${t_{S_1} + \frac{t_{T}}{2}}$.  This suggests that we should increase the size of the hexagonal face of the tile as much as possible.  Notice, $t_{S_2}$ does not appear in the equation~\ref{t_ov_2}.

For the above formulation, we assumed that the tiles are compute bound.  We need a mechanism to first prune the tile size design space and restrict it to only consider compute bound tiles and then use the above cost functions to further reduce the search space.

\section{The effect of the hyper-threading}
Our results in ~\cite{prajapati2017simple} suggest that we should revisit the ``conventional wisdom'' that says that an optimal strategy of a tiling is to choose the ``largest possible tile size that fits'' i.e., its memory footprint matches the available capacity.  First of all, this falls into the trap that it precludes overlapping of computation and communication (the ``hyperthreading effect''). But this can be avoided by explicitly accounting for hyperthreading.  Indeed, our GPU platforms preclude such large size by disallowing the data footprint of a thread block to exceed \emph{half} the shared memory capacity.

Thus, the hyperthreading-adjusted ``conventional wisdom'' would still seek to maximize tile volume subject to the half-capacity constraint---the best strategy is the largest tile volume for the given footprint.  Our model and experimental data suggests otherwise---an even higher hyperthreading factor is turning out to yield the best performance.  We still don't know why, and it is subject to investigation.

% Roofline
% Hong-Kung Lower Bounds
% Sorting Accelerator bottleneck
% how to do bottleneck analysis using time/energy/area/codesign
%
% Show Problem Formulation and Design Space using models

%\include{chap7-Other}
% Other / Meta Properties
% 1. Performance Portability - preserving performance while moving from one architecture to another
% 2. Human Insights 
%     a. Reduce Overheads - Large Hexagons - Prune Tile Size Design Space
%     b. Communication/Computation overlap

%\include{Timeline}
% chapter? Proposed Project Timeline
% Work Plan and Implications

\chapter{Related Work}
\label{chap:related}
%%%%%%%%%%%%%%%%%%%%%%%%%%%%%%%%%%%%%%%%%%%%%%%%%%%%%%%%%%%%%%%%

% + Literature Review / Related work / Background
% + Talk about performance models and its uses.

Our research draws upon prior work in the following distinct areas.

\section{Stencil Computations and Code Generation}

At the algorithmic level, most stencil applications are \emph{compute bound} in the sense that the ratio of the total number of \emph{operations} to the total number of \emph{memory locations} touched can always be made ``sufficiently large'' because it is an asymptotically increasing value.  We may expect that such codes can be optimized to achieve very high performance relative to machine peak.  However, naive implementations turn out to be memory-bound.  Therefore, many authors seek to exploit data locality for these programs~\cite{kamil2010,liu2009,bandishti12}.  One successful technique is called \emph{time tiling}~\cite{Wolf91tiling,uday08pldi,wonnacott99,wonnacott02,frigo-etal-focs99,frigo-strumpen-ics05,StShPa_11CATS,bandishti12}, an advanced form of loop tiling~\cite{Wol87,Wolf91tiling,Xue00tiling}.  Time tiling first partitions the whole computation space into tiles extending in all dimensions, and then optionally executes these tiles in a so called ``45 degree wavefront'' fashion.  We assume, like most of the work in the literature, that \emph{dense} stencil programs are \emph{compute bound} after time tiling. However, due to the intricate structure of time tiled code, writing it by hand is challenging.  Automatic code generation, is an attractive solution, and has been an active research topic.
% These techniques are applicable whether the target platform is a multi-core
% CPU or an accelerator like a GPU.

For iterative stencils a large set of optimizing code generation strategies have been proposed.  Pochoir~\cite{Tang2011Pochoir} is a CPU-only code generator for stencil computations that exploits reuse along the time dimension by recursively dividing the computation in trapezoids.  Diamond tiling~\cite{Bandishti12diamond_tiling}, Hybrid-hexagonal tiling~\cite{grosser-etal-GPUhextile-CGO2014}, and Overtile~\cite{Holewinski2012Overtile} are all tiling strategies that allow to exploit reuse along the time dimension, while ensuring a balanced amount of coarse-grained parallelism throughout the computation.  While the former has only been evaluated on CPU systems, the last two tiling schemes have been implemented to target GPUs.  Overtile uses redundant computation whereas hybrid-hexagonal tiling uses the hexagonal tiles to avoid the need for redundant computation and the increased shared memory that would otherwise be required to store temporary values.  Another time tiling strategy has been proposed with 3.5D blocking by Nguyen et.~al~\cite{Nguyen2010.3.5DBlockingForCPUandGPU}, who manually implemented kernels that use two dimensional space tiling plus streaming along one space dimension with tiling along the time dimension to target both CPUs and GPUs. A slightly orthogonal stencil optimization has been proposed by Henretty et.~al, who use data-layout transformations to avoid redundant non-aligned vector loads on CPU platforms.

\section{Performance Modeling}

%There has been much work on time modeling and performance optimization.  
All of the previously discussed frameworks either come with their own auto-tuning framework or require auto tuning to derive optimal tile sizes.  For stencil graphs, which are directed acyclic graphs (DAGs) of non-iterated stencil kernels, various DSLs compilers have been proposed. Halide~\cite{Ragan-Kelley2013Halide} and Stella~\cite{Gysi2015Stella} are two DSLs from the context of image processing and weather modeling that separate the specification of the stencil computation from the execution schedule, which allows for the specification of platform specific execution strategies derived either by platform experts or automatic tuning.  Both DSLs support various hardware targets, including CPUs and GPUs. Polymage~\cite{Mullapudi2015Polymage} also provides a stencil graph DSL---this time for CPUs only---but pairs it with an analytical performance model for the automatic computation of optimal tile size and fusion choices.  With MODESTO~\cite{Gysi2015Modesto} an analytical performance model has been proposed that allows to model multiple cache levels and fusion strategies for both GPUs and CPUs as they arise in the context of Stella.

For stencil GPU code generation strategies that use redundant computations in combination with ghost zones, an analytical performance model has been proposed~\cite{Meng2009GhostZhonePerfModel} that allows to automatically derive ``optimal'' code generation parameters.  Yotov et.~al~\cite{Yotov2003} showed already more than ten years ago that an analytical performance model for matrix multiplication kernels allows to generate code that is performance-wise competitive to empirically tuned code generated by ATLAS~\cite{ClintWhaley20013}, but at this point no stencil computations have been considered.  Shirako~et~al.~\cite{Shirako2012} use cache models to derive lower and upper bounds on cache traffic, which they use to bound the search space of empirical tile-size tuning.  Their work does not consider any GPU specific properties, such as shared memory sizes and their impact on the available parallelism.  In contrast to tools for tuning, Hong and Kim~\cite{Hong2009} present a precise GPU performance model which shares many of the GPU parameters we use.  It is highly accurate, low level, and requires analyzing the PTX assembly code.  For stencil GPU code generation strategies that use redundant computations in combination with ghost zones an analytical performance model has been proposed~\cite{Meng2009GhostZhonePerfModel} that allows to automatically derive ``optimal'' code generation parameters.  Patus~\cite{Christen11Patus} provides an auto-tuning environment for stencil computations which can target CPU and GPU hardware. It does not use software managed memories and also does not consider any time tiling strategies.  

Renganarayana et al.~\cite{Renganarayana-2008-PPT} identifies positivity as a common property shared by the parameters used by tile size selection methods and show that the property can be used to derive efficient and scalable tile size selection frameworks. 

\section{Chip Reverse Engineering and Area Modeling}

Chip area modeling can be formally considered a branch of semiconductor reverse engineering, which is a well researched subject area.  Torrence et.~al.~\cite{Torrence:2009} gives an overview of the various techniques used for chip reverse engineering.  The packaged chips are usually decapped and the wafer die within is photographed layer by layer.  The layers are exposed in the reverse order after physical or chemical exfoliation. {\em Degate}~\cite{degate}, for example, is a well known open source software that can help in analyzing die photographs layer by layer.  The reverse engineering process can be coarse-grained to identify just the functional macro-blocks.  Sometimes, the process can be very fine-grained, in order to identify standard-cell interconnections, and hence, actual logic-gate netlists.  Degate is often used in association with catalogs of known standard cell gate layouts, such as those compiled by {\em Silicon Zoo}~\cite{SiliconZoo}.  Courbon et.~al.~\cite{Courbon:2016} provides a case study of how a modern flash memory chip can be reverse engineered using targeted scanning electron microscope imagery.  For chip area modeling, one is only interested in the relatively easier task of demarcating the interesting functional blocks within the die.

\section{Energy Modeling}
\label{sec:energyrelated}

GPU power/energy model is a very active area: a recent survey article on the topic~\cite{Mittal2014} cites almost~150 references.  We only discuss the relevant work here.  The model we present complements Mittal and Vetter~\cite{Mittal2014} by enabling us to find the optimal parameters (i.e., tile sizes) for the energy efficient execution of stencil like programs.
Hong and Kim~\cite{Hong10} present a GPU power model to predict the number of optimal GPU cores to achieve the peak memory bandwidth for a kernel.  An analytical model is used to predict the execution time~\cite{Hong09} which has enabled prediction of the power consumption statically.  However, they have predicted the minimum number of cores required for a program to achieve the peak memory bandwidth of GPU.  While this approach may work for memory bandwidth bound programs, it is unlikely to produce better results for compute-bound programs like tiled stencil computations.  Our model is much simpler, because our model does not depend on warp and thread level parameters and number of PTX instructions.

Nagasaka et al~\cite{Nagasaka10} model GPU power of kernels using performance counters.  Lim et al.~\cite{Lim14:GPUMcPAT}, GPUWattch~\cite{Leng13:GPUWattch} and GPUSimPow~\cite{Lucas13:GPUSimPow} are simulation based power models. McPAT~\cite{Li09:McPAT} is the basis for Lim et al~\cite{Lim14:GPUMcPAT} and GPUWattch~\cite{Leng13:GPUWattch} uses GPGPUSim~\cite{Bakhoda09:GPGPUSim} to simulate execution time.  Simulation and performance counters-based models require execution (or simulation) of the program to predict the power consumption.  Therefore, these models are not feasible solutions when it requires to take decisions at compile time to determine optimal software parameters.  We do need to run some micro-benchmarks to find the energy parameters that our model use.  In contrast, we run our micro-benchmarks only to determine parameters of a GPU architecture, while the power consumption can be predicted for a given program statically
%%\footnote{Similar to Hong and Kim~\cite{Hong10}, a
%%time model which is capable of predicting time statically~\cite{Hong09} an be
%%used to model the execution time of the program.} 
without running the program.

There are studies~\cite{Ren11,Ren12} focused on reducing the energy for both CPU and GPU by balancing the load among CPU and GPU. Our study is only focused on modeling the energy consumption of GPUs. There are studies~\cite{Ren11,Ren12} focused on reducing the energy for both CPU and GPU.  The models are used to determine how to balance the load among CPU and GPU, so that it reduces the overall energy consumption.  Our study is only focused on modeling the energy consumption of GPUs.

\section{Codesign}

% Add a paragraph here about the standard work on application codesign.

Application codesign is a well established discipline and has seen active research for well over two decades~\cite{Prakash-Parker1992, Wolf-1997, Chatha-Vemuri-Magellan-CODES2001, Dick-Jha-TCAD2006-mogac, Teich-proceedings2012}.  The essential idea is to start with a program (or a program representation, say in the form of a CFDG---Control Data Flow Graph) and then map it to an abstract hardware description, often represented as a graph of operators and storage elements.  The challenge that makes codesign significantly harder than compilation is that the hardware is not fixed, but is also to be synthesized.  Most systems involve a search over a design space of feasible solutions, and various techniques are used to solve this optimization problem: tabu search and simulated annealing~\cite{eles-etal-todaes97, Erbas-etal-TEC2006}, integer linear programming~\cite{Niemann-Marwedel1997}.

There is some recent work on accurately modeling the design space, especially for regular, or \emph{affine control} programs~\cite{Pouchet-etalFPGA2013, Zuo-etal-ICCAD2013, Zuo-etal-CODES+ISSS2013}.  However, all current approaches solve the optimization problem for a single program at a time.  To the best of our knowledge, no one has previously considered the \emph{generalized application codesign} problem, seeking a solution for a \emph{suite of programs}.

There are multiple publications on codesign related to exascale computing, but they focus on different aspects.  For instance, Dosanji et al.~\cite{Dosanjh:2014:EDS:2562354.2562814} focus on methodological aspects of exploring the design space, including architectural testbeds, choice of mini-applications to represent applications codes, and tools.  The ExaSAT framework \cite{doi:10.1177/1094342014568690} was developed to automatically extract parameterized performance models from source code using compiler analysis techniques.  Performance analysis techniques and tools targeting exascale and codesign are discussed in \cite{b1282de6ab8a4f1a8a962cb40310bc7f}. 

Kuck et al.~\cite{Kuck2012codesign, Kuck-keynote2013codesign} analyze and model program hot-spots. They develop computational capacity models and propose an approach for the HW/SW codesign of computer systems.  The hardware/software measurements of computational capacity (based on bandwidth usage) and power consumption (based on hardware counters) are used to find optimal solutions to various codesign problems and to evaluate codesign trade-offs.  Their models are theoretical and are illustrated by numerical examples.  They do not validate their models using real hardware. 

%\section{Bottleneckology}
%+ Roofline
%+ Hong-Kung Lower Bounds

% + Literature Review / Related work / Background
% + Talk about performance models and its uses.

\chapter{Conclusions}
\label{chap:conclusions}
%%%%%%%%%%%%%%%%%%%%%%%%%%%%%%%%%%%%%%%%%%%%%%%%%%%%%%%%%%%%%%%%

%CONTRIBUTION to KNOWLEDGE

Our work contributes to knowledge in following ways:

\paragraph{The unified view of the polyhedral design landscape}We put together all the parameters to be considered for performance optimization in a single \emph{\textbf{unified landscape}} (Chapter~\ref{chap:approach}).  The landscape (shown in Figure~\ref{fig:landscape}) considers the program, architecture, and compiler parameters and combines them with various cost metrics.  This view lets us identify the pockets of domain specificity and allows us to study performance improvement across all cost metrics.

\paragraph{Analytical Models}We develop \emph{\textbf{analytical cost models}} (Chapter~\ref{chap:costmodels}) for execution time, energy, memory access, and silicon area of a chip.  Our models are reasonably accurate and help predict the associated cost. We argue that these models can be used to break the HPC application performance improvement cycle.

\paragraph{Mathematical Optimization Approach}We formulate mathematical optimization problems to address some of the challenges of exascale.  We show how these optimizations can be used for \emph{\textbf{performance tuning}} (Chapter~\ref{chap:autotuning}) and \emph{\textbf{accelerator codesign}} (Chapter~\ref{chap:codesign}).  For GPGPU stencil computations and polyhedral code generator, we illustrate a proof of concept~\cite{prajapati2017techreport} and present a novel optimization approach to accelerator codesign.

\section{Limitations of our approach}
Our approach is limited to the narrow area of domain specific applications, polyhedral model and GPU-like programmable hardware accelerators.  The approach can, however, be extended to other set of \emph{programs/architectures/transformations} by identifying other domain specific regions in the design landscape.  More work is needed in order to extend our approach across different regions of domain specificity.

%Our models are not accurate for all codes and can be improved.  Currently, our execution time model do not model threads and registers.  Number of threads is known to impact performance of codes.  The number of registers available per thread is limited and hence

\section{Open Questions}
Among the many different uses of the analytical cost models, they can be further explored to answer important performance related questions.  We list some of them below:
\begin{itemize}
\item Using analytical execution time and energy models we can find out
(i) What happens when input parameters change?
(ii) What happens when different number of processors is used?
(iii) What is the largest possible problem size on a given architecture?
(iv) When does the efficiency drop?
\item Silicon area models can be used for the following:
(i) Chip area prediction.
(ii) Generate all possible configurations for a give die area.
(iii) Calculate the cost of different configurations.
(iv) Study the trade-offs.
(v) System tuning.
\item Performance tuning related questions such as
(i) Sensitivity of problem size to tile size.
(ii) Sensitivity of optimal tile size for different codes.
(iii) Reconfirm the folklore : whether optimizing for time is equal to optimizing for energy?
(iv) Reasons for the poor performance.
can be answered using cost models.
\end{itemize}

The answer to these questions become helpful in two situations.  One, for \emph{performance portability} while moving from one architecture to another.  Second, obtaining interesting insights to recognize promising areas for future research.

\backmatter % starts unnumbered supplementary material
%%%%%%%%%%%%%%%%%%%%%%%%%%%%%%%%%%%%%%%%%%%%%%%%%%%%%%%%%%%%%%%%

% Bibliography
%%%%%%%%%%%%%%%%%%%%%%%%%%%%%%%%%%%%%%%%%%%%%%%%%%%%%%%%%%%%%%%%

% unsorted BibTeX style
% check here for more:  https://www.sharelatex.com/learn/Bibtex_bibliography_styles
\bibliographystyle{unsrt}
\bibliography{prelims-references,bib/CHACK,bib/references,bib/codesign,bib/bib4,bib/parallelizing,bib/energy}
%,bib/paper} %change sample to the name of your .tex file, e.g., thesis

\appendix % starts the appendices
%%%%%%%%%%%%%%%%%%%%%%%%%%%%%%%%%%%%%%%%%%%%%%%%%%%%%%%%%%%%%%%%

%\include{appendixA}

% PPoPP
%\include{PPoPP/PPoPP-main}

% HPPAC
%\include{HPPAC/Energy-main}

% CODESIGN
%\include{Codesign/Codesign-main}

% Have a nice day!
%%%%%%%%%%%%%%%%%%%%%%%%%%%%%%%%%%%%%%%%%%%%%%%%%%%%%%%%%%%%%%%%
\end{document}